\documentclass[12pt]{JHEP3}

\usepackage{amscd,amsmath,amssymb,amsfonts,xspace,mathrsfs,bbm}
\usepackage{graphicx}
\usepackage{fancyhdr}
\usepackage{rotating}
\usepackage{psfrag}
\usepackage{epsfig}
\usepackage{url}
\usepackage{bbold}
\usepackage{undertilde}
\usepackage{color}
\let\normalcolor\relax

\allowdisplaybreaks
\newcounter{tabl}
\newcommand{\be}{\begin{equation}}
\newcommand{\ee}{\end{equation}}
\def\ba{\begin{eqnarray}}
\def\ea{\end{eqnarray}}
\newcommand{\beq}{\begin{eqnarray}}
\newcommand{\eeq}{\end{eqnarray}}
\newcommand{\bea}[2]{\be\label{#2}\begin{array}{#1}}
\newcommand{\eea}{\end{array}\ee}

\def\det{\,{\rm det}\, }
\def\diag{{\rm diag}}

\def\({\left(}
\def\){\right)}
\def\[{\left[}
\def\]{\right]}
\def\p{\partial}
\newcommand{\nn}{\nonumber}
\newcommand{\de}{\mathrm{d}}

\def\11{1\!\! 1}

\def\hf{\frac{1}{2}}
\def\hft{{\textstyle\frac{1}{2}}}

\def\a{\alpha}
\def\b{\beta}

\def\eps{\varepsilon}

\def\l{\lambda}
\def\m{\mu}
\def\n{\nu}

\def\r{\rho}

\def\s{\sigma}

\def\G{\Gamma}

\def\L{\Lambda}

\usepackage{ulem}
\def\na{\nabla}
\def\om{\omega }
\def\Om{\Omega }
\def\w{\wedge}

\DeclareFontFamily{U} {MnSymbolC}{}

\DeclareFontShape{U}{MnSymbolC}{m}{n}{
  <-6> MnSymbolC5
  <6-7> MnSymbolC6
  <7-8> MnSymbolC7
  <8-9> MnSymbolC8
  <9-10> MnSymbolC9
  <10-12> MnSymbolC5 
  <12-> MnSymbolC12}{}
\DeclareFontShape{U}{MnSymbolC}{b}{n}{
  <-6> MnSymbolC-Bold5
  <6-7> MnSymbolC-Bold6
  <7-8> MnSymbolC-Bold7
  <8-9> MnSymbolC-Bold8
  <9-10> MnSymbolC-Bold9
  <10-12> MnSymbolC-Bold10
  <12-> MnSymbolC-Bold12}{}

\DeclareSymbolFont{MnSyC} {U} {MnSymbolC}{m}{n}
\DeclareMathSymbol{\Istar}{\mathrel}{MnSyC}{128}

   \def\CA {{\cal A}}
   \def\CB {{\cal B}}
   \def\CC {{\cal C}}
   \def\CD {{\cal D}}
   
   \def\CF {{\cal F}}
   \def\CG {{\cal G}}
   \def\CH {{\cal H}}

   \def\CM {{\cal M}}

   \def\CS {{\cal S}}

   \def\CV {{\cal V}}

\newcommand{\tX}{\lefteqn{\smash{\mathop{\vphantom{<}}\limits^{\;\sim}}}X}
\newcommand{\tE}{\lefteqn{\smash{\mathop{\vphantom{<}}\limits^{\;\sim}}}E}

\newcommand{\tP}{\lefteqn{\smash{\mathop{\vphantom{<}}\limits^{\;\sim}}}P}

\newcommand{\Et}{\lefteqn{\smash{\mathop{\vphantom{\Bigl(}}\limits_{\sim}
\atop \ }}E}
\newcommand{\Pt}{\lefteqn{\smash{\mathop{\vphantom{\Bigl(}}\limits_{\sim}
\atop \ }}P}

\newcommand{\eet}{\lefteqn{\smash{\mathop{\vphantom{\biggl(}}\limits_{\sim}
\atop \ }}\,e}

\newcommand{\Nt}{\lefteqn{\smash{\mathop{\vphantom{\Bigl(}}\limits_{\sim}
\atop \ }}N}

\newcommand{\teps}{\tilde{\eps}}
\newcommand{\epst}{\lefteqn{\smash{\mathop{\vphantom{\Bigl(}}\limits_{\!\scriptstyle{\sim}}\atop \ }}\eps}

\def\hCG{\hat\CG}

\def\hCH{\hat\CH}
\def\hCV{\hat\CV}
\def\hN{\hat N}

\def\bee{\bar e}

\newcommand{\Ref}[1]{\eqref{#1}}

\newcommand{\f}{\frac}

\def\xc{{\rm x}}
\def\yc{{\rm y}}

\newcommand{\bfN}{\boldsymbol{N}}

\newcommand{\bfCH}{\boldsymbol{\CH}}
\newcommand{\bfpsi}{\boldsymbol{\psi}}

\newcommand{\hbfN}{\hat \bfN}

\def\omp{\omega_+}
\def\omm{\omega_-}
\def\ompm{\omega_\pm}

\def\ompi#1{\omega_{+,#1}}
\def\ommi#1{\omega_{-,#1}}
\def\ompmi#1{\omega_{\pm,#1}}

\def\Xp{X_+}
\def\Xm{X_-}

\def\tXp{\tX_+}
\def\tXm{\tX_-}
\def\tXpm{\tX_\pm}
\def\tXmp{\tX_\mp}
\def\tXpi#1{\tX_{+,#1}}
\def\tXmi#1{\tX_{-,#1}}

\def\tXmpi#1{\tX_{\mp,#1}}

\def\eep{e_+}
\def\eem{e_-}
\def\eepm{e_\pm}
\def\eemp{e_\mp}
\def\eepi#1{e_{+,#1}}
\def\eemi#1{e_{-,#1}}
\def\eepmi#1{e_{\pm,#1}}
\def\eempi#1{e_{\mp,#1}}

\def\fpi#1{f_{+,#1}}
\def\fmi#1{f_{-,#1}}
\def\fpmi#1{f_{\pm,#1}}

\def\tPp{\tP_+}
\def\tPm{\tP_-}
\def\tPpm{\tP_\pm}

\def\tPpi#1{\tP_{+,#1}}
\def\tPmi#1{\tP_{-,#1}}
\def\tPpmi#1{\tP_{\pm,#1}}

\def\Ptpmi#1{\Pt_{\pm,#1}}

\def\Ep{E_+}
\def\Em{E_-}
\def\Epm{E_\pm}

\def\Epmi#1{E_{\pm,#1}}

\def\tEpmi#1{\tE_{\pm,#1}}

\def\Etpmi#1{\Et_{\pm,#1}}

\def\chip{\chi_+}
\def\chim{\chi_-}
\def\chipm{\chi_\pm}

\def\chipmi#1{\chi_{\pm,#1}}

\def\etapmi#1{\eta_{\pm,#1}}

\def\Np{N_+}
\def\Nm{N_-}
\def\Npm{N_\pm}

\def\Ntp{\Nt_+}
\def\Ntm{\Nt_-}
\def\Ntpm{\Nt_\pm}
\def\Ntmp{\Nt_\mp}

\def\Omp{\Om_+}
\def\Omm{\Om_-}
\def\Ompm{\Om_\pm}

\def\rpm{r_\pm}

\def\rpmi#1{r_{\pm,#1}}

\def\qp{q_+}
\def\qm{q_-}
\def\qpm{q_\pm}
\def\qmp{q_\mp}
\def\qpi#1{q_{+,#1}}

\def\qpmi#1{q_{\pm,#1}}

\def\ddp{d_+}
\def\ddm{d_-}
\def\ddpm{d_\pm}

\def\PhiD{\mathscr{D}}

\def\PhiH{\mathscr{H}}

\def\mhD{\hat\PhiD^{\rm int}}
\def\mHp{\PhiH_+^{\rm int}}
\def\mHm{\PhiH_-^{\rm int}}
\def\mHpm{\PhiH_\pm^{\rm int}}

\newcommand{\gaux}{{\mathfrak g}}
\newcommand{\eaux}{{\mathfrak e}}

\title{Bi-gravity with a single graviton}

\author{
Sergei Alexandrov$^{1}$ and Simone Speziale$^2$
\\
$^1${\it Laboratoire Charles Coulomb (L2C), Universit\'e de Montpellier,
CNRS, F-34095, Montpellier, France}\\
$^2${\it Centre de Physique Th\'{e}orique, CNRS-UMR 7332, Luminy Case 907, 13288 Marseille, France}

\vspace*{2mm} {\tt e-mail:
\email{sergey.alexandrov@umontpellier.fr}, \email{simone.speziale@gmail.com}
}

\vspace*{-3mm}

}

\abstract{We analyze a bi-gravity model based on the first order formalism, having as fundamental variables two tetrads but only one Lorentz connection.
We show that on a large class of backgrounds its linearization agrees with general relativity. 
At the non-linear level, additional degrees of freedom appear, and we reveal the mechanism hiding them around the special backgrounds.
We further argue that they do not contain a massive graviton, nor the Boulware-Deser ghost.
The model thus propagates only one graviton, whereas the nature of the additional degrees of freedom remains to be investigated.
We also present a foliation-preserving deformation of the model,
which keeps all symmetries except time diffeomorphisms and has
three degrees of freedom.
}

\begin{document}

\section{Introduction}

Bi-gravity and massive gravity provide modifications of general relativity (GR) that have given rise to much theoretical
and phenomenological discussions in the literature, motivated by finding alternatives to $\L$CDM
or by exploring potential quantum gravity models. Interest spiked after the identification
of ghost-free interactions \cite{deRham:2010ik,deRham:2010kj,deRham:2011rn},
which eliminated a first obvious obstruction in the viability of these models --- the so-called Boulware-Deser (BD) ghost \cite{Boulware:1973my}
which typically appears in massive gravity models at non-linear level.
Although the ghost-free interactions are somewhat cumbersome-looking in the metric language,
they are very natural in tetrad variables, where they appear as
the only polynomials generalizing the cosmological constant term
which can be written using differential forms
\cite{Hinterbichler:2012cn} (see also \cite{Chamseddine:2011mu}).\footnote{There is also a similar formulation using self-dual 2-forms
\cite{Alexandrov:2012yv},
albeit only for the symmetric one of the 3 possible ghost-free interaction terms. See also \cite{Alexandrov:2008fs,Speziale:2010cf,Beke:2011mu}.}

Nevertheless, there are other difficulties which plague these models,
including a-causal propagation \cite{Gruzinov:2011sq,deFromont:2013iwa,Deser:2013eua,Deser:2013uy}
and reappearing of ghost instabilities around cosmological homogeneous solutions
\cite{DeFelice:2012mx,DeFelice:2013bxa}. See \cite{Babichev:2013usa,Mukohyama:2015vsa} for recent reviews.
In the bi-gravity context \cite{Hassan:2011tf,Hassan:2011zd},
these difficulties are induced by the degrees of freedom of the second, massive graviton.
This raises the question of whether it is possible to find infrared modifications of GR in
which eventual additional degrees of freedom do {\it not} behave like a massive graviton.\footnote{Bi-gravity models
are not the only ones haunted by a massive graviton.
For instance, Stelle's renormalizable theory of higher derivative gravity \cite{Stelle:1976gc}
also propagates it, and this mode spoils unitarity of the theory. This said, quantum corrections could restore unitarity
by making the massive mode unstable, see e.g. \cite{Donoghue:2018izj}. }

At the same time, the beauty and simplicity of the tetrad formulation of the ghost-free interactions
calls for being part of the story.
But how can one avoid the appearance of the second graviton in the presence of the two tetrads
used to build these interactions?
Consider the following reasoning.

The other three known interactions of nature are all carried by a connection field
taking values in the appropriate Lie algebra.
While Einstein's general relativity uses the metric as fundamental variable, it is known that it can be reformulated
as a gauge theory, using a tetrad and a connection valued in the Lie algebra of the Lorentz group.
The connection is however algebraically determined by the tetrad,
and therefore does not have a fundamental dynamical role.
But regarding the connection as the carrier of gravitational interactions suggests that
even if one takes {\it several} tetrads, but charged under the {\it same} Lorentz connection,
one would describe just one graviton.

With such a motivation in mind, in this paper we analyze a
model that implements this idea. Defined by two Einstein-Cartan actions plus the polynomial 
interactions mentioned above, see \eqref{totalaction} below, it can be viewed either as a simple modification
of the first order action of GR, or as a variation on the theme of ghost-free bi-gravity.
It takes as fundamental variables two tetrads, but only one connection.
Although the connection is still algebraically determined by the tetrads by
the equations of motion, it is a complicated function thereof
which does not coincide with the usual Levi-Civita solution.
The resulting dynamics is thus quite different from both GR and bi-gravity theories,
and so is the identification of the physical degrees of freedom.

A version of the model considered here has already appeared in the literature \cite{deRham:2015rxa},
but with only one tetrad dynamical and the second one frozen,
thus regarded as a modification of massive gravity rather than bi-gravity.
It was found that the model propagates a massive graviton plus some additional degrees of freedom,
although the analysis was not conclusive about their number.
In related papers \cite{deRham:2013tfa,deRham:2015cha}, it was also shown that any attempt to modify the kinetic term of GR
(and our modification can indeed be considered as a modification of the kinetic term)
would introduce a BD ghost, since at least one of the second class constraints removing it would be lost.
As we show in this paper, the situation with both tetrads dynamical and just one connection is quite different.

Our first result is that the linearization of the model around a large class of backgrounds --- including the doubly-flat one ---
contains a single massless spin-2 particle which obeys the linearized Einstein's equations, in agreement with the diffeomorphism
invariance of the action. This implies that this modified theory of gravity
is indistinguishable from GR in this linear regime, despite involving two tetrads.
This is in a striking difference with standard bi-gravity where the linear spectrum always contains a massive graviton.

Next, we perform the canonical analysis of the model at non-linear level, finding that additional degrees of freedom do appear.
Specifically, the model has 8 physical degrees of freedom in total.
This is deceivingly the same number as in old bi-gravity models containing a massless graviton (2 degrees of freedom),
a massive graviton (5) and the scalar BD ghost (1), however in our case
the behavior of the degrees of freedom and hence their interpretation are different. 
First of all, the 5+1 additional degrees of freedom of bi-gravity models show up at linear level around the doubly-flat background,
whereas as stated above, in our model they are hidden around this particular background,
as well as around a much larger class.\footnote{The hiding property,
that we consider here interesting for phenomenological applications, is on the other hand often frowned upon because
it implies a strong coupling for perturbative expansions {\it near} the hiding backgrounds \cite{ArkaniHamed:2002sp}.
However, this strong coupling problem does not prevent us from considering such models as
effective field theories provided the kinetic terms generated by loop effects do not lead to any instability.
Furthermore, the negative viewpoint seems to stem, in our opinion, from the prejudice that field theories,
especially their quantum versions, should exist perturbatively.
The existence of non-perturbative approaches such as loop quantum gravity makes us inclined to go beyond this prejudice.}
In addition, the origin of the additional modes is very different from
the origin of the massive graviton in bi-gravity, and more similar to the additional modes discussed in \cite{deRham:2015rxa}.
Hence, we argue that there is {\it no} massive graviton in the spectrum.
Furthermore, we also show that the spectrum is free from the BD ghost because
both constraints removing it in the usual ghost-free bi-gravity are still present.
These are the two desired features that we expected from our model.
At the same time, the precise geometric nature of the additional degrees of freedom remains to be investigated.

In the end of the paper we also consider a modification of our model obtained by imposing an additional constraint,
which can also be viewed as a restriction of the original model to a particular sector of the phase space.
The additional constraint introduces a preferred foliation and breaks time diffeomorphisms, but
is consistent with all other symmetries
and leads to a drastic reduction of degrees of freedom. We show that the modified model propagates only 3 degrees of freedom,
similarly to various known examples of foliation-preserving modifications of GR \cite{Jacobson:2008aj,Horava:2009uw,Barvinsky:2019agh}.

The organization of the paper is as follows.
In the next section we present the model and discuss its general features.
In section \ref{sec-linear} we study its linearization,
first around the doubly flat background and then around arbitrary conformally related tetrads.
Next, in section \ref{sec-kinetic} we explain the results of linearization from the analysis of the kinetic terms.
In section \ref{sec-canan} we provide the complete canonical analysis.
A special attention is paid to the case without potential terms of the two tetrads when the model is shown to
possess two additional gauge symmetries. In section \ref{sec-sector} we present a foliation-preserving
modification of the model. Finally, section \ref{sec-disc} is devoted to conclusions and discussion.
A few appendices contain details of calculations and some useful formulae.

Our conventions are such that the internal space indices $I,J=0,\dots,3$ are raised and lowered by means of the flat Minkowski metric
$\eta_{IJ}=\diag(-,+,+,+)$. The Levi-Civita symbol with flat indices is normalized as $\eps_{0123}=1$. On the other hand,
for the antisymmetric tensor density with spacetime indices we use $\teps^{0123}=1$ and 
$\teps^{0abc}=\teps^{abc}$ where $a,b,\dots$ label spatial directions.
The symmetrization and anti-symmetrization of indices are denoted by $(\cdot\, \cdot)$ and $[\cdot\,\cdot]$, respectively,
and include the factors of 1/2.

\section{The model}
\label{sec-model}

General relativity can be described in the first order formalism, using tetrads and a Lorentz connection as independent fields,
by the Einstein-Cartan action \cite{Hehl:1976kj}\footnote{In units $8\pi G=1$, convenient to avoid numerous factors of 2 in the canonical analysis.
Sometimes this action is also called Hilbert-Palatini or tetrad Palatini formulation,
in reference to Palatini's first order formulation in metric variables, see \cite{Peldan:1993hi}.
}
\be
S_{\rm 0}[e,\omega]=\frac14\int \eps_{IJKL}e^I\wedge e^J \wedge F^{KL}(\omega),
\label{action-Pal}
\ee
where $F^{IJ}(\omega)=\, \de \omega^{IJ}+{\omega^I}_K \wedge \omega^{KJ}$. 
The action is invariant under diffeomorphisms as well as internal Lorentz transformations.
In bi-gravity models written in this formalism \cite{Hinterbichler:2012cn,Alexandrov:2013rxa}
one takes the sum of two actions $S_0$ with independent tetrads and connections, plus an interaction term:
\be
S_{\scriptsize{\mbox{bi-g}}}[\eep,\eem,\omp,\omm]= S_{0}[\eep,\omp]+S_{\rm 0}[\eem,\omm]
+S_{\rm int}[\eep,\eem].
\label{totalaction-bg}
\ee
For generic interactions one finds 8 degrees of freedom including the scalar BD ghost. Ghost-free models propagate only 7 degrees of freedom,
corresponding to a massless and a massive graviton around the doubly flat spacetime, and are characterized by the following five interaction terms:
\be
S_{\rm int}[\eep,\eem]
= -\int \eps_{I_1I_2I_3I_4}\sum_{k=0}^4 \frac{\beta_k}{k!(4-k)!} \(\mathop{\wedge}\limits_{i=1}^k \eep^{I_i} \)\wedge\(\mathop{\wedge}\limits_{i=k+1}^4 \eem^{I_i} \).
\label{Sint}
\ee
The mass-dimension-2 parameters are two cosmological constants $\beta_0$ and $\beta_4$,
and three coupling constants $\beta_{1,2,3}$ determining in turn the mass of the second spin-2 particle
appearing in the spectrum.
It is a beautiful consequence of working with tetrads that
the ghost-free interactions are the simplest ones that can be written down, and the only ones
that can be written exclusively using differential forms.

The model considered in this paper is obtained by keeping in the above action two different tetrads, but a single Lorentz connection.
This leads to the following action
\be
S[\eep,\eem,\omega]=\frac14\int \eps_{IJKL}\( \eep^I\wedge \eep^J+\eem^I\wedge \eem^J\) \wedge F^{KL}(\omega)
+S_{\rm int}[\eep,\eem],
\label{totalaction}
\ee
where the interaction term $S_{\rm int}$ is taken to be as in \Ref{Sint}.
Notice that we assumed the `Newton's constants' of each sector to be positive,
and reabsorbed them into a rescaling of the tetrads. The signs are relevant for the physical interpretation of the model,
and our analysis can easily be generalized to arbitrary signs.
One could also consider the mixed term where the curvature is multiplied by $\eep^I\wedge \eem^J$, but it can be removed by
a linear redefinition of the tetrads.
Up to these redefinitions, \Ref{totalaction} is the most general action constructed from one Lorentz connection and two tetrads
regarded as differential forms,
which is linear in the curvature and invariant under diffeomorphism and Lorentz gauge transformations including parity.

\subsection{Gauge symmetries}
\label{subsec-gauge}

The action \eqref{totalaction} is clearly invariant under `diagonal' local Lorentz transformations and diffeomorphisms,
namely those acting
in the same way on both tetrads. For the usual ghost-free bi-gravity \Ref{totalaction-bg},
it makes sense to speak also about local Lorentz transformations and diffeomorphisms acting independently on each set of variables.
They are symmetries of the first two terms in \eqref{totalaction-bg}, but not of the interaction term \eqref{Sint}
generating the mass for one of the two gravitons.
In contrast, in the model \eqref{totalaction} the `off-diagonal' symmetries are broken already by the kinematical first term.
As was noticed in \cite{deRham:2015rxa} (where the same action \Ref{totalaction} was considered,
but with only one tetrad dynamical and the second fixed), due to this fact one loses the symmetricity constraint
\be
\eta_{IJ}\eepi{\mu}^I\eemi{\nu}^J=\eta_{IJ}\eemi{\mu}^I\eepi{\nu}^J,
\label{symcon}
\ee
responsible for the equivalence between the metric and tetrad formulations of bi-gravity.
The reason is the following. In the standard bi-gravity case, only the term $S_{\rm int}$ contributes
to the difference of the variations of the action \eqref{totalaction-bg}
with respect to Lorentz gauge degrees of freedom in the two sectors, since the first two terms are individually invariant.
The resulting equations \eqref{symcon} are purely algebraic, and allow one to exclude the corresponding `off-diagonal' degrees
of freedom even though the `off-diagonal' Lorentz transformations are not a symmetry of the full action.
With the new action \eqref{totalaction}, the kinetic term of the action does contribute to the same (difference of)
variations, so that the resulting equation includes the curvature and turns out to be dynamical, instead of being a constraint like \eqref{symcon}.
As a consequence, a metric implementation  of the same idea explored here, namely an action  $S(g_{+},g_-,\G)$
given by two Einstein-Hilbert actions sharing the same affine connection,
plus the metric version of the ghost-free interactions, will in general give a different dynamics.

In the particular case where all parameters $\beta_k$ are set to zero, i.e. in the absence of all interaction terms
including the cosmological constant ones,
the action \eqref{totalaction} turns out to have two additional gauge symmetries.
At the infinitesimal level, they act by
\begin{subequations}\label{addsym}
\beq
\CS_1:\ &&
\eepmi{\mu}^I\to \eepmi{\mu}^I\pm \eps\eempi{\mu}^I,
\label{addgsym1}
\\
\CS_2:\ &&
\sqrt{\eepm}\eepmi{I}^\mu\to \sqrt{\eepm}\(\eepmi{I}^\mu\pm \eps \eemp\eempi{I}^\mu\),
\label{addgsym2}
\eeq
\end{subequations}
where $\eps$ is a transformation parameter.
The first symmetry is evident by inspection of the action. The second becomes transparent if one uses
\be
\frac14\,\teps^{\mu\nu\rho\sigma}\eps_{IJKL}e_\rho^K e_\sigma^L=e e^\mu_{[I} e^\nu_{J]}
\ee
to rewrite the first term in \eqref{totalaction} as
\be
\int\de^4 \xc\, \(\eep \eepi{I}^\mu \eepi{J}^\nu+\eem \eemi{I}^\mu \eemi{J}^\nu\)F_{\mu\nu}^{IJ}(\omega).
\ee
The appearance of these additional symmetries was quite surprising to us. In fact,
we initially inferred their existence from the constraint analysis, and only afterwards identified them at the covariant level.
As we show below in section~\ref{subsec-noint}, the canonical realization of the second symmetry is particularly non-trivial.
Interestingly, the canonical analysis will show that the presence of the additional symmetries does not change the number of degrees of freedom.

\subsection{Equations of motion}
\label{subsec-coneq}

An important difference of the model \eqref{totalaction} with respect to the standard bi-gravity \eqref{totalaction-bg}
is the new form of the connection equation. 
Instead of the usual Cartan equation $d_\om e^I=0$ 
(here $d_{\om}$ is the covariant exterior derivative with respect to the Lorentz connection $\omega^{IJ}$)
with the unique Levi-Civita solution $\om^{IJ}(e)$, one has
\be
d_{\om} B^{IJ}=0, \qquad
B^{IJ}\equiv\eep^{I}\wedge \eep^{J}+ \eem^{I}\wedge \eem^{J}.
\label{Cartaneq2}
\ee
This equation is still algebraic with respect to the connection and can be rewritten as
\be\label{domB}
B^{[IK}\wedge {\omega^{J]}}_K=\eep^{[I}\wedge d\eep^{J]}+ \eem^{[I}\wedge d\eem^{J]},
\ee
however it is clear that its solution is in general not a Levi-Civita connection, neither the sum of two Levi-Civita connections.

Furthermore, since the 2-form $B^{IJ}$ is not simple (i.e. it is not a wedge product of two 1-forms),
it is quite involved to define the inverse of the algebraic operator acting on the connection
and find the general solution. Although in principle such general solution is known \cite{Beke:2011mu}
(see also \cite{Cuesta:2008xu}), it is a non-linear expression in $B^{IJ}$ constructed using the Urbantke metrics defined from a general 2-form.
This solution does not appear to significantly simplify in the special case \eqref{Cartaneq2} of our current interest,
and we leave investigations of this approach to future work.

Instead, we note that the connection equation can be easily solved for any pair of conformally equivalent tetrads, say
\be
\eepm^I = \Ompm e^I.
\label{lin-confeq}
\ee
In this case the 2-form $B^{IJ}$ is simple,
\be
B^{IJ}=e^{2\Phi}\, e^I\wedge e^J,
\qquad
\Phi=\hf\, \log \(\Omp^2+\Omm^2\),
\label{defBPhi}
\ee
and the connection is given by the Levi-Civita connection of $e^I$
plus a contorsion piece determined by $\Phi$,
\be
\om_\mu^{IJ} = \om_\mu^{IJ}(e)+2 e^{[I}_\mu e^{J]\nu}\p_\nu\Phi.
\label{solom-confeq}
\ee
One can define then an `effective' tetrad
\be
\eaux^I\equiv e^\Phi e^I
\label{efftet}
\ee
for which the solution \eqref{solom-confeq} reduces to the standard Levi-Civita connection
\be
\om_\mu^{IJ} = \om_\mu^{IJ}(\eaux),
\label{solcon-eff}
\ee
as follows from the transformation properties of the Levi-Civita connection under conformal transformations.
As we will see below, it is the effective metric constructed from $\eaux^I$
\be
\gaux_{\mu\nu}=\eta_{IJ}\eaux_\mu^I\eaux_\nu^J=\(\Omp^2+\Omm^2\)\eta_{IJ}e_\mu^Ie_\nu^J,
\label{effmet}
\ee
that acts as background for the massless graviton.

As for the tetrad equations, they give two sets at first sight similar to the ordinary Einstein equations in tetrad language:
\begin{align}\label{tetradeqs}
\f14 \,\eps_{IJKL} \teps^{\m\n\r\s} e_{\pm,\nu}^J F^{KL}_{\r\s}(\om) = U_{\pm I}^\m(\b_k;e_+,e_-),
\end{align}
where the right-hand side comes from the straightforward variations of \Ref{Sint}.
However because the connection is in general not Levi-Civita, its curvature is not equivalent to the Riemann tensor,
and the left-hand side does not reproduce the Einstein tensor.

\section{Linearization}
\label{sec-linear}

To get a first understanding of the dynamics of the model and its degrees of freedom, in this section we study the linearization of
\eqref{totalaction} around a given background.
Following the standard treatment of perturbation theory in tetrad variables (see e.g. \cite{Deser:1974cy}), we define
\be
\eepmi{\mu}^I=\bee_{\pm,\mu}^I+\fpmi{\mu}^I,
\qquad
\omega_\mu^{IJ}=\bar\om^{IJ}_\m+ w_\mu^{IJ},
\label{expandvar}
\ee
where $\bee_{\pm,\mu}^I$ and $\bar\om^{IJ}_\m$ describe the (on-shell) background,
whereas $\fpmi{\mu}^I$ and $w_\mu^{IJ}$ are the infinitesimal perturbations.
It is also useful to define the tensorial variables
\be\label{deffmn}
f_{\pm,\m\n}\equiv \eta_{IJ}f^I_{\pm,\m}\bar e_{\pm,\n}^J,
\ee
which are related to the metric perturbations.
Plugging the expansion \eqref{expandvar} into the action and expanding to the quadratic order, one gets the linearized theory.
Analyzing the resulting linearized field equations, one can then elucidate the physical meaning of the perturbations.

\subsection{Doubly-flat background}
\label{subsec-flat}

The first case we study is the doubly-flat background, which in ghost-free bi-gravity allows one to provide a physical
interpretation to all propagating degrees of freedom. It is defined by
\be
\bar e_{\pm,\mu}^I=\delta_\mu^I,
\qquad
\bar\omega_\mu^{IJ}=0.
\ee
Expanding the action at first order, it is immediate to see that the doubly-flat background
is a solution of the equations of motion only provided the parameters satisfy
the following two conditions
\be
\begin{split}
\lambda_+\equiv&\,\beta_1+3\beta_2+3\beta_3+\beta_4=0,
\\
\lambda_-\equiv&\,\beta_0+3\beta_1+3\beta_2+\beta_3=0,
\end{split}
\label{condbeta}
\ee
corresponding to the absence of an effective cosmological constant.
These conditions allow to exclude two of the parameters $\beta_k$, for instance, $\beta_1$ and $\beta_3$.

Introducing the `diagonal' and `off-diagonal' combinations of the tetrad fluctuations
\be
f_{\mu\nu}\equiv \fpi{\mu\nu}+\fmi{\mu\nu},
\qquad
b_{\mu\nu}\equiv \fpi{\mu\nu}-\fmi{\mu\nu},
\label{defcombfluct-flat}
\ee
the action expanded to the second order reads
\be
S_{(2)}= \int\de^4 \xc\Bigl[{f_\rho}^\rho\p_{\mu}w_\nu{}^{\mu\nu}+{f_\mu}^\rho\p_{\nu}w_\rho{}^{\mu\nu}-{f_\mu}^\rho\p_{\rho}w_\nu{}^{\mu\nu}
+2\,\eta_{\rho\sigma} w_{[\mu}{}^{\mu\rho}w_{\nu]}{}^{\sigma\nu}-\frac{\beta}{2}\,{b_\mu}^{[\mu}{b_\nu}^{\nu]}
\Bigr],
\ee
where all indices are raised and lowered with the flat Minkowski metric, and can be converted from spacetime to internal space indices and back
using the flat tetrad $\delta_\mu^I$. 
Here $\beta\equiv \hf\(\beta_0-2\beta_2+\beta_4\)$ is the only parameter which remains in the linearized theory,
and we observe that the `off-diagonal' variables decouple:
for $\beta\ne 0$ they are fixed by their equations of motion to zero,
whereas in the case of vanishing $\beta$ they remain undetermined describing pure gauge degrees of freedom,
corresponding to an additional shift gauge symmetry of the linearized theory.

The equation obtained by varying $w_\mu{}^{\a\b}$ is linear and algebraic, and has solution 
\be
w_\m{}^{\a\b} =\hf\, \eta^{\n[\a}\eta^{\b]\r}\(\p_{\r}s_{\m\n} + \f12\p_\m a_{\n\r}\)
\label{solom}
\ee
in terms of the symmetric $s_{\mu\nu}\equiv 2f_{(\mu\nu)}$ and antisymmetric  $a_{\mu\nu}\equiv 2 f_{[\mu\nu]}$ parts of the `diagonal' fluctuations.
This expression is proportional to the Levi-Civita connection for fluctuations $f_{\m\n}$ around the effective metric $2\eta_{\mu\nu}$,
see appendix~\ref{ap-linear} for details.\footnote{The factor of 2
can be understood from \eqref{effmet} and setting $\Om_\pm=1$.
Had we used the effective tetrad $\eaux_\mu^I=\sqrt{2}\delta_\mu^I$ to convert indices, we would obtain
the right proportionality factor 1/4 in agreement with the doubly-flat limit of the more general formula \eqref{genres-w} derived below.}

Inserting the solution into the equation obtained by varying $f_{\mu\nu}$,
one finds that the antisymmetric part $a_{\mu\nu}$
drops out and remains unrestricted, whereas the symmetric part satisfies the following differential equation
\be
G^{(1)}_{\m\n}[\eta;s]
= -\hf\Bigl(\p^2 s_{\mu\nu} -\p_\mu\p_\rho {s_\nu}^\rho-\p_\nu\p_\rho {s_\mu}^\rho+\p_\mu\p_\nu {s_\rho}^\rho
+\eta_{\mu\nu}\(\p_\rho\p_\sigma s^{\rho\sigma}-\p^2 {s_\rho}^\rho\)\Bigr)=0.
\label{Einstlinflat}
\ee
The second order differential operator appearing in \eqref{Einstlinflat}
is nothing but the linearized Einstein tensor $G_{\m\n}$ on a flat background.

This analysis shows that the only propagating mode visible in the linear spectrum around the doubly-flat background
is the massless spin-2 graviton, described by $s_{\m\n}$.
The internal Lorentz fields $a_{\m\n}$ are pure gauge, as in usual tetrad gravity.
The crucial result is that the `off-diagonal' field $b_{\m\n}$, which carries the
massive graviton mode in usual bi-gravity theories, is not dynamical in this linearization.

\subsection{Conformally-related backgrounds}
\label{subsec-conf}

One may think that the reduction to Einstein's gravity at the linearized level and absence of additional propagating modes are a
consequence of the high symmetry of the doubly-flat background. As we now show, these features persist in more general conformally-equivalent backgrounds,
namely
\be
\bar e_\pm^I = \Ompm \bar e^I.
\label{lin-confflat}
\ee
Let us first study the equations of motion that the background has to satisfy.

The equation for the connection can be solved as anticipated in Section~\ref{subsec-coneq},
so that
$\bar\omega_\mu^{IJ}$ is
the Levi-Civita connection evaluated on the effective tetrad $\bar\eaux^I=e^\Phi \bar e^I$, with $\Phi$ defined as in \eqref{defBPhi}.
Substituting this solution into the variation of the action with respect to the tetrads, one finds the two equations
\be
G_{\m\n}(\bar\gaux) = -e^{-2\Phi}\frac{\lambda_\pm}{\Ompm} \,\bar\gaux_{\m\n},
\label{Einsteq-pm}
\ee
where $G_{\m\n}(\bar\gaux)$ is the Einstein tensor evaluated on the effective metric
$\bar\gaux_{\mu\nu}=\eta_{IJ}\bar\eaux_\mu^I\bar\eaux_\nu^J$, and
\be
\begin{split}
\l_+ \equiv&\, \b_1\Om_-^3+3\b_2\Om_-^2\Om_++3\b_3\Om_+^2\Om_-+ \b_4\Om_+^3,
\\
\l_- \equiv &\, \b_3\Om_+^3+3\b_2\Om_+^2\Om_-+3\b_1\Om_-^2\Om_++ \b_0\Om_-^3
\end{split}
\label{deflampm}
\ee
generalizing $\lambda_\pm$ introduced in \eqref{condbeta}.
Taking the difference of the two equations \eqref{Einsteq-pm}, one finds a constraint on $\Ompm$
\be
\Omm\lambda_+=\Omp\lambda_-.
\label{constrOmpm}
\ee
This algebraic constraint can be solved for generic values of the coupling constants.
There is no restriction on them as in the doubly-flat case, hence this perturbative expansion tests generic properties of our action.
In addition, the Bianchi identity requires that the function multiplying the metric on the right hand side of \eqref{Einsteq-pm}
be a constant, which can be identified with the effective cosmological constant $\Lambda$.
Thus, we have a stronger constraint
\be
e^{-2\Phi}\frac{\lambda_+}{\Omp}=e^{-2\Phi}\frac{\lambda_-}{\Omm}\equiv\Lambda =\mbox{const}.
\label{defLam}
\ee
Finally, the two conformal factors must be such that the effective metric $\bar\gaux_{\mu\nu}$
satisfies the standard Einstein equations with cosmological constant $\Lambda$.

Next, we study the equations on the perturbations $w_\mu^{IJ}$ and $\fpmi{\mu\nu}$,
which again can be conveniently decomposed in terms of `diagonal' and `off-diagonal' combinations \eqref{defcombfluct-flat},
with the former split into its symmetric $s_{\mu\nu}$ and antisymmetric $a_{\mu\nu}$ parts.
The analysis is a bit longer than in the doubly-flat case, we thus present here an overview of the results and report the details in Appendix \ref{ap-linear}.
\begin{itemize}
\item The solution for the connection is given by a covariant version of \Ref{solom},
\be
w_\m{}^{\a\b} = \bar\gaux^{\n[\a}\bar\gaux^{\b]\r}\(\bar\na_\r s_{\m\n}+\f12\bar\na_\m a_{\n\r}\),
\label{genres-w}
\ee
where $\bar\na$ is the Levi-Civita connection of $\bar\gaux_{\mu\nu}$.
Crucially, this solution is again a linearization of the Levi-Civita connection for the `diagonal' perturbations $f_{\mu\nu}$
around the effective background  $\bar\gaux_{\mu\nu}$.

\item
The symmetric diagonal perturbations $s_{\mu\nu}$ satisfy
the linearized Einstein equations on the background described by the effective metric $\bar\gaux_{\mu\nu}$ with cosmological constant \Ref{defLam}, namely
\be
G^{(1)}_{\m\n}[\bar\gaux;s] + \L s_{\m\n} = 0.
\label{linEinst}
\ee

\item The antisymmetric perturbations $a_{\mu\nu}$ are pure gauge.

\item
The `off-diagonal' perturbations $b_{\mu\nu}$ are non-dynamical: they are fixed in terms of the diagonal fields by an algebraic equation,
\be
\(\delta_\mu^\rho\delta_\nu^\sigma-W (\Om_+^2+\Om_-^2)\bar C_\m{}^\s{}_\n{}^\r\)b_{\r\s}
=\(B\,\delta_\mu^\rho\delta_\nu^\sigma-W (\Om_+^2-\Om_-^2)\bar C_\m{}^\s{}_\n{}^\r\)f_{\r\s},
\label{eq-bf}
\ee
where $\bar C_{\mu\nu\rho\sigma}$ is the Weyl tensor of the effective background metric,
and $B$ and $W$ are functions of the background and parameters $\b_k$ defined in \eqref{defW}.

\end{itemize}

We conclude that at linear order the model propagates only one massless graviton around any
background described by two conformally related tetrads. The main difference with respect to the doubly flat case is that
the `off-diagonal' perturbations do not have to vanish, but are fixed in terms of the solution of the linearized Einstein equations.

We also remark that all solutions of the linearized GR are included in our model,
provided one can split the background metric as in \eqref{effmet}
so that the two functions $\Ompm$ satisfy the constraint \eqref{defLam}.

\section{Kinetic terms}
\label{sec-kinetic}

In this section we switch to a canonical analysis to explain
why the linear spectrum on conformally related backgrounds
contains only one massless graviton and no additional degrees of freedom.
To that end, we introduce a $3+1$ decomposition of spacetime and analyze the kinetic terms.
We parametrize the tetrads as in the familiar ADM decomposition \cite{Arnowitt:1960es,Arnowitt:1962hi},
using a tilde over or under the fields to indicate spatial density of positive or negative weight:
\be
\eepm^I=\(\Ntpm \tXpm^I+\Npm^a \eepmi{a}^I\)\de t + \eepmi{a}^I\de x^a.
\label{dec-tetrad}
\ee
Here $\eepmi{a}^I$ and $\tXpm^I$ satisfy
\be
\eta_{IJ}\tXpm^I\eepmi{a}^J=0,
\qquad
\eta_{IJ}\tXpm^I \tXpm^J=-\qpm,
\qquad
\eta_{IJ} \eepmi{a}^I \eepmi{b}^J=\qpmi{ab},
\label{propvectors}
\ee
and  $\qpmi{ab}$ are the two induced metrics on the spatial slice, $\qpm$ are their determinants.
An explicit solution of these relations is given by\footnote{
Notice also that in the single metric case the norm of the vector $\chi^i$ conveniently controls the nature
of the foliation, from the standard space-like one to time-like \cite{Alexandrov:2005ar}
and null \cite{Alexandrov:2014rta}.}
\be
\begin{split}
\tXpm^I=\,\Epm
\(1,\chipm^i\),
\qquad
\eepmi{a}^I=\,\(\Epmi{a}^j\chipmi{j},\Epmi{a}^i\).
\end{split}
\label{parameX}
\ee
Here the indices $i,j=1,2,3$ are used to label spatial directions in the internal space,
$\Epmi{a}^i$ are spatial triads and $E_{\pm}$ their determinants.
We will denote their inverses by $\Epmi{i}^a$, and the densities $\tEpmi{i}^a=\Epm\Epmi{i}^a$.

The kinetic term of \Ref{totalaction} following from this decomposition is
\be
S_{\rm kin}=\int \de t\, \de^3\xc \(\tPpi{IJ}^a+\tPmi{IJ}^a\)\p_t \omega_a^{IJ},
\label{Skin}
\ee
where we introduced
\be
\tPpmi{IJ}^a\equiv \frac14\, \teps^{abc} \eps_{IJKL} \eepmi{b}^K \eepmi{c}^L
=
\left\{
\begin{array}{ccl}
\tfrac12\, \tEpmi{i}^a && \mbox{for } [IJ]=[0i],
\\
\tEpmi{[i}^a\chipmi{j]}&\quad & \mbox{for } [IJ]=[ij].
\end{array}
\right.
\label{deftP}
\ee
Since the momentum conjugate to $\omega_a^{IJ}$ is not an elementary field,
the structure of the kinetic term should be further disentangled.
This can be achieved through a change of variables
which generalizes the one used to bring the kinetic term to the canonical form in the case of one
tetrad (see \cite{Alexandrov:1998cu} and \eqref{decomp-ompm} in the last appendix),
\begin{subequations}\label{decomp-omEchi}
\begin{align}
& \tEpmi{i}^a=\hf\,\tE^a_j(\delta_i^j\pm d^j_i),
&& \chipm^i=\chi^i-\zeta_{[i} d^j_{j]}\pm \zeta^i,
\\
& \omega_a^{0i}=\eta_a^i-\omega_a^{ij}\chi_j+\omega_a^{ij}\zeta_{[j}d^k_{k]}-\omega_a^{kj}\zeta_j d^i_k,
&& \omega_a^{ij}=\eps^{ijk}r_{kl}\Et_a^l +\Et_a^{[i}\omega^{j]}.
\label{decomp-om-sym}
\end{align}
\end{subequations}
Here the first two relations define `diagonal' ($\tE^a_i$, $\chi^i$) and `off-diagonal' ($d_i^j$, $\zeta^i$) components of the tetrad
(analogously to, but in a more complicated way than in \eqref{defcombfluct-flat}),
whereas the last two trade the 18 components of the connection $\omega_a^{IJ}$ for $\eta_a^i$,
the symmetric matrix $r_{ij}$ and the internal spatial vector $\omega^i$.
Using these new variables allows us to write the kinetic term of the action in the following form:
\be
S_{\rm kin}=\int \de t\, \de^3\xc \Bigl[\tE^a_i\p_t \eta_a^i+\chi_i\p_t\omega^i
+\eps^{ikl}d^j_k\zeta_l\,\p_t r_{ij}\Bigr].
\label{Skin4}
\ee
The first two terms are in the canonical form and familiar from the single tetrad \cite{Alexandrov:1998cu}
and standard bi-gravity cases \cite{Alexandrov:2013rxa}.
On the other hand, the momentum conjugate to $r_{ij}$ is still not an elementary field. While we were not able to find
a further change of variables which brings it to the canonical form,
this expression for the kinetic term is sufficient to draw some definite conclusions.

In the case of GR in the first order formalism, there is only one tetrad, the fields $d_i^j$ and $\zeta^i$
describing the `off-diagonal' sector are absent, and so is the non-canonical last term in \eqref{Skin4}.
Hence the six components $r_{ij}$ of the connection have vanishing momenta and are non-dynamical fields.
In our case this is not true anymore. The momenta conjugate to $r_{ij}$ are non-vanishing and are given by
\be
\pi^{ij}=\eps^{(ikl}d^{j)}_k\zeta_l.
\label{defpi}
\ee
As a result, one expects that the model propagates up to six additional degrees of freedom,
in agreement with the results of \cite{deRham:2015rxa}.

Let us consider however the linearization around any background with conformally related tetrads \eqref{lin-confflat}.
The crucial observation is that for such backgrounds
\be
\bar d_i^j\sim \delta^i_j,
\qquad
\bar\zeta^i=0.
\label{difbackgr}
\ee
Therefore, at linearized level the momenta \eqref{defpi} are again vanishing!
This fact explains why the linear spectrum around these backgrounds contains only a massless graviton.
It also shows that the hiding mechanism at play is based on the non-linear dependence on the `off-diagonal' variables 
of the momenta of the additional degrees of freedom.

\section{Canonical analysis}
\label{sec-canan}

In this section we perform the complete canonical analysis of the model \eqref{totalaction} and find the number of propagating
degrees of freedom on generic backgrounds.
Notice first that the action of our model can be equivalently written as 
the action of the standard ghost-free bi-gravity \eqref{totalaction-bg} constrained to have equal connections
\be
S[\eep,\eem,\omp,\omm,\upsilon]=S_{\scriptsize{\mbox{bi-g}}}[\eep,\eem,\omp,\omm]+\int\de^4 \xc\, \upsilon^\mu_{IJ}(\ompi{\mu}^{IJ}-\ommi{\mu}^{IJ}),
\label{totalaction-alt}
\ee
where $\upsilon^\mu_{IJ}$ is a Lagrange multiplier imposing the constraint.
Integrating out $\upsilon^\mu_{IJ}$, one recovers the original action \eqref{totalaction}.
Choosing \eqref{totalaction-alt} as the starting point allows to make the canonical analysis similar to the one of the ghost-free bi-gravity,
to see their parallels  and differences.

\subsection{Primary constraints}
\label{subsec-decomp}

For the purpose of canonical analysis, it is convenient to treat the time components of the connection
directly as Lagrange multipliers.
To be able to do this, we need to integrate out $\upsilon^0_{IJ}$ in the action \eqref{totalaction-alt} which imposes $\ompi{0}^{IJ}=\ommi{0}^{IJ}$,
but leaves the other components independent, thus making one step back towards the original formulation \eqref{totalaction}.
Then decomposing the tetrads as in \eqref{dec-tetrad}, the action can be rewritten in the following canonical form
(cf. \cite{Peldan:1993hi,Alexandrov:2000jw,Alexandrov:2013rxa})
\be
\begin{split}
S=&\, \int \de t\, \de^3\xc\Bigl[\tPpi{IJ}^a\p_t \ompi{a}^{IJ}+\tPmi{IJ}^a\p_t \ommi{a}^{IJ}
+ \upsilon^a_{IJ}(\ompi{a}^{IJ}-\ommi{a}^{IJ})
+\omega_0^{IJ}\CG_{IJ}
\\
&\,
+\Np^a\CV_{+,a}+\Nm^a\CV_{-,a}+\Ntp\CH_+ +\Ntm\CH_-\Bigr],
\end{split}
\label{decompaction}
\ee
where the momenta $\tPpmi{IJ}^a$ are given by \eqref{deftP}.
The explicit expressions of
$\CG_{IJ}$, $\CV_{\pm,a}$ and $\CH_\pm$ can be found in \eqref{primconstr}.
They form the set of primary constraints, together with
\be
\psi_a^{IJ}= \ompi{a}^{IJ}-\ommi{a}^{IJ}.
\label{constr-psi}
\ee
One must also take into account that not all components of $\tPpmi{IJ}^a$
are independent: as for usual gravity in tetrad variables, we have 6 simplicity constraints per sector,
\be
\phi_\pm^{ab}=\hf\,\eps^{IJKL}\tPpmi{IJ}^a\tPpmi{KL}^b.
\label{conphi}
\ee
Thus, in total one has $6+2\times 4+18+2\times 6=44$ primary constraints. They are defined on a $72$-dimensional phase space
parametrized by $\tPpmi{IJ}^a$ and $\ompmi{a}^{IJ}$, with symplectic structure
\be
\{\ompmi{a}^{IJ}(\xc), \tPpmi{KL}^b(\yc)\} =  \delta_a^b\delta^{IJ}_{KL}\delta(\xc,\yc).
\label{canPB}
\ee

Before we proceed with the study of the constraint algebra, it is convenient to replace $\CV_{\pm,a}$ by their linear combinations.
Namely, we define
\be
\begin{split}
\CD_a\equiv &\, \CV_{+,a}+\CV_{-,a}+\hf\,(\ompi{a}^{IJ}+\ommi{a}^{IJ})\CG_{IJ}
\\
&\, +\psi_a^{IJ}\[\hf\(\p_b\tPpi{IJ}^b-\p_b\tPmi{IJ}^b\)+{\ompi{bI}}^{K}\tPpi{KJ}^b-{\ommi{bI}}^{K}\tPmi{KJ}^b \]
\\
=&\, \p_b\(\tPpi{IJ}^b\ompi{a}^{IJ}+\tPmi{IJ}^b\ommi{a}^{IJ}\)-\tPpi{IJ}^b\p_a\ompi{b}^{IJ}-\tPmi{IJ}^b\p_a\ommi{b}^{IJ},
\\
\hCV_a\equiv&\, \hf\,(\CV_{+,a}-\CV_{-,a}).
\end{split}
\label{constr-diff-alt}
\ee
The advantage of this redefinition is that $\CD_a$ generates the standard spatial diffeomorphisms in both sectors.\footnote{Note that the interaction
term $S_{\rm int}$ does not contribute to $\CD_a$.}
Similarly, the Gauss constraint $\CG_{IJ}$ generates the local Lorentz transformations. It is easy to see that both these constraints commute weakly
with all other constraints in agreement with their first class nature as generators of gauge transformations.

Let us now consider the simplicity constraints \eqref{conphi} together with the constraints \eqref{constr-psi}.
Their Poisson brackets are non-vanishing so that one can expect that they form second class pairs.
But there are 18 constraints $\psi_a^{IJ}$ and only 12 $\phi_\pm^{ab}$ so that there should exist 6 combinations of $\psi_a^{IJ}$ which commute with
all simplicity constraints. One can check that this is indeed the case and the following constraints have vanishing commutation
relations with $\phi_\pm^{ab}$:
\begin{subequations}\label{psiX}
\beq
\psi_a&= &\tXp^I\psi_{a,IJ}\tXm^J,
\label{comp-psia}
\\
\psi_\pm &= & -2\tPpmi{IK}^a\tXpm^K \psi_a^{IJ}\tXmp^J=\qpm \qpm^{ab}\eepmi{b}^I\psi_a^{IJ}\tXmp^J,
\label{comp-psipm}
\\
\psi_0 &= &
\hf\,\teps^{abc}\eps_{IJKL} \eepi{b}^K\eemi{c}^L \psi_a^{IJ},
\label{comp-psi0}
\eeq
\end{subequations}
where $\qpm^{ab}$ are the inverses of the induced metrics defined in \eqref{propvectors}.
We denote the set of these constraints by $\psi_X$ with $X$ running over 6 values.

As a result, we remain with the set of constraints $\psi_X$, $\hCV_a$ and $\CH_\pm$
for which we should study the stability conditions. This problem is very similar to what one has in the ghost-free bi-gravity \cite{Alexandrov:2013rxa}:
the only difference is that there, instead of $\psi_X$, one has the off-diagonal Gauss $\hCG_{IJ}$ given by the difference of
the Gauss constraints in the two sectors.
Let us recall that in the case of bi-gravity the stabilization of $\hCG_{IJ}$ gives rise to three secondary constraints $\CS^a$
which coincide with the spatial components of the symmetricity conditions \eqref{symcon}.
Then $\hCG_{IJ}$ form second class pairs with $\hCV_a$ and $\CS^a$.
At the same time, the stability of the two Hamiltonian constraints $\CH_\pm$ also
generates the secondary constraint \cite[Eq.(3.14)]{Alexandrov:2013rxa}
\be
\Psi= \CM^a_{IJ} \(\ompi{a}^{IJ}-\ommi{a}^{IJ}\),
\label{Psiexpr}
\ee
where
\be
\CM^a_{IJ}=
\eps_{IJKL}\teps^{abc}\(\beta_1 \eemi{b}^K \eemi{c}^L+2\beta_2 \eepi{b}^K\eemi{c}^L+\beta_3 \eepi{b}^K\eepi{c}^L\),
\ee
which in turn forms a second class pair with a linear combination of $\CH_\pm$, whereas another combination remains first class.
It is trivial to check that this constraint structure leads to 14-dimensional phase space, i.e. to 7 degrees of freedom of one massless
and one massive gravitons, whereas the constraint $\Psi$ \eqref{Psiexpr} is nothing but the constraint removing the BD ghost.

In our case the constraint structure is certainly going to be different since we already know that
the symmetricity conditions do not arise in our model.
On the other hand, it is interesting that the constraint removing the BD ghost is still imposed.
Indeed, the constraint \eqref{Psiexpr} is simply a linear combination of our constraints $\psi_a^{IJ}$.
Thus, both primary and secondary constraints responsible for the absence of the BD ghost in bi-gravity appear now as primary constraints,
which allows us to conclude that our model is free from the BD ghost.

\subsection{Absence of the interaction terms}
\label{subsec-noint}

Before treating the general case, it is useful to consider the case where all parameters $\beta_k$ are taken to vanish,
i.e. the interaction term $S_{\rm int}$ is not included. As was noticed in section \ref{sec-model}, then the model possesses
two additional gauge symmetries \eqref{addsym}.
Therefore, the first natural question is: what are the first class constraints generating them?
Since the symmetry transformations do not affect the connections,
it is clear that the generators must be linear combinations of the constraints \eqref{psiX}.
Comparing their commutators with the canonical variables against the transformations given in appendix \ref{ap-sym},
one can check that the generator of $\CS_1$ coincides with $\psi_0$, whereas the generator of $\CS_2$ is given by
\be
\psi=\hN^a\psi_a+\Ntp\psi'_+ +\Ntm\psi'_-,
\label{genS2}
\ee
where $\hN^a=\Np^a-\Nm^a$ is the Lagrange multiplier of $\hCV_a$.

These identifications make sense only if the corresponding constraints $\psi_0$ and $\psi$ are first class, i.e.
commute with all other constraints. Potentially, the only non-vanishing commutators can be with $\hCV_a$ and $\CH_\pm$.
We provide their explicit expressions in appendix \ref{ap-com} and prove that $\psi_0$ indeed weakly commutes with all constraints.

The situation with $\psi$ is more complicated. First of all, it is already a bit unusual that one has to consider a quantity
which is not really a function on the phase space since it also involves Lagrange multipliers.
In fact, this is analogous to the generator of time diffeomorphisms in any generally covariant theory which
is known to coincide with the {\it full} Hamiltonian (see, e.g. \cite{Reisenberger:1995xh,Alexandrov:2001wt}),
i.e. it is the linear combination of first class constraints with coefficients given by the corresponding Lagrange multipliers.
But here there is an additional complication: the constraints entering \eqref{genS2} are not first class!
Indeed, the matrix of commutators of $\psi_{A'}=(\psi_a,\psi_+,\psi_-)$ with $\CH_A=(\hCV_a,\CH_+,\CH_-)$ has the following structure
\be
\{\psi_{A'}(\xc),\CH_B(\yc)\}\approx M_{A'B}\delta(\xc,\yc),
\qquad
M_{A'B}=\(
\begin{array}{ccc}
\CA_{ab} & \CB^+_a &\CB^-_a
\\
- \CB^+_b & 0 & -\CC
\\
-\CB^-_b & \CC & 0
\end{array}\),
\label{matrixM}
\ee
where $\CA_{ab}$, $\CB^\pm_a$ and $\CC$ are defined in \eqref{defABC}.
However, it turns out that the vanishing of all commutators is not necessary for $\psi$ to be a symmetry of the action!
To see why this is so, let us write the action in the canonical form as
\be
S=\int \de t\(p^i\p_t q_i+n^\alpha G_\alpha\),
\label{canonaction}
\ee
where $q_i$, $p^i$ are our canonical variables, $G_\alpha$ is the set of all primary constraints and $n^\alpha$ are their Lagrange multipliers.
Furthermore, we know that
\be
\begin{split}
\{\psi_{A'},G_\alpha\}=&\, C_{A\alpha}^\gamma G_\gamma,  \quad \alpha\ne B
\\
\{\psi_{A'},\CH_B\}=&\, C_{A'B}^\gamma G_\gamma+M_{A'B},
\end{split}
\label{algGG}
\ee
and under the action generated by $\eps^\alpha G_\alpha$, the canonical variables and
the Lagrange multipliers transform as
\be
\begin{split}
&\, \delta q_i= \{\eps^\alpha G_\alpha, q_i\},
\qquad
\delta p^i=\{\eps^\alpha G_\alpha, p^i\},
\\
&\,\qquad\quad\
\delta n^\alpha= \p_t\eps^\alpha- C^\alpha_{\beta\gamma}\eps^\beta n^\gamma.
\end{split}
\label{transLM}
\ee
For the generator $\psi$ \eqref{genS2}, one must take $\eps^\alpha=\eps N^A$ for $\alpha=A'$, where $N^A=(\hN^a,\Ntp,\Ntm)$, and zero otherwise.
Then applying the transformation \eqref{transLM} to the action \eqref{canonaction} and using the constraint algebra \eqref{algGG},
it is easy to see that the variation vanishes provided the matrix $M_{A'B}$ is antisymmetric.
Remarkably, this is indeed the case for the matrix \eqref{matrixM}.

Let us now turn to the stability conditions for the remaining constraints, i.e $\psi_{A'}$ and $\CH_A$.
It is clear that they reduce to the equations
\be
M_{A'B}N^B=0,
\qquad
\upsilon^{B'} M_{B'A}=\cdots
\label{stab-Hpsi}
\ee
where $\upsilon^{B'}$ are Lagrange multipliers for $\psi_X$ and
the dots denote contributions from commutators of $\CH_A$ with all other constraints.
From the first equation, one finds
\be
\Ntp=\CC^{-1}\CB^-_a\hN^a,
\qquad
\Ntm=-\CC^{-1}\CB^+_a\hN^a,
\ee
and
\be
\CM_{ab}\hN^b=0,
\qquad
\CM_{ab}=\CA_{ab}+\CC^{-1}\(\CB^+_a\CB^-_b-\CB^-_a\CB^+_b\).
\label{eqMhN}
\ee
Since $\CM_{ab}$ is a $3\times 3$ antisymmetric matrix, it has a vanishing determinant and
the equations \eqref{eqMhN} fix only 2 of the 3 components of $\hN^a$ as
\be
\hN^a=\Nt\teps^{abc}\CM_{bc}.
\ee
This is consistent with the fact that one combination of the constraints $\CH_{A}$ is first class,
whereas 4 remaining constraints are second class.
In the following, it will be convenient to use notations $\bfN^A$ for the functions on the phase space
determining the Lagrange multipliers $N^A$ up to the factor of $\Nt$ so that we have $N^A=\Nt\bfN^A$.

It is clear that the second equation in \eqref{stab-Hpsi} has a similar solution. More precisely, one obtains
$\upsilon^{A'}=\Nt\bfN^A+\cdots$ where the dots correspond to terms proportional to the Lagrange multipliers of the constraints contributing
to the r.h.s. of that equation (i.e. $\phi_\pm^{ab}$ and those $\psi_a^{IJ}$ which form with $\phi_\pm^{ab}$ second class pairs).
Importantly, one Lagrange multiplier remains unfixed implying that one combination of $\psi_{A'}$ is first class, whereas 4 others are second class.

Explicitly, the combinations of the constraints which are first class can be written as
\be
\bfCH=\hbfN^a\hCV_a+\bfN_+\CH_++\bfN_-\CH_- +\cdots,
\qquad
\bfpsi=\hbfN^a\psi_a+\bfN_+\psi'_++\bfN_-\psi'_- .
\label{totHampsiS}
\ee
where this time the dots denote terms proportional to $\phi_\pm^{ab}$ and $\psi_a^{IJ}$ which can be fixed from the stability equations of these constraints.
The first class constraint $\bfpsi$ can be thought also as the generator of the gauge symmetry $\CS_2$.
However, this is true only on mass shell: this constraint is proportional to the generator \eqref{genS2} provided the Lagrange multipliers are
set to their values fixed by the stability conditions.

We conclude that in the absence of the interaction term the 72-dimensional phase space carries
12 first class constraints $\CG_{IJ}$, $\CD_a$, $\bfCH$, $\psi_0$, $\bfpsi$.
and 32 second class constraints comprising $\phi_\pm^{ab}$, 16 components of $\psi_a^{IJ}$ and 4 constraints out of $(\hCV_a,\CH_+,\CH_-)$.
This leaves $72-2\times 12-32=16$ dimensional phase space, i.e. 8 degrees of freedom.
Two of them correspond to the massless graviton, whereas the remaining six can be viewed as the components of the connection $r_{ij}$
becoming dynamical on general background, as suggested by \eqref{Skin4}.

\subsection{Inclusion of the interaction terms}
\label{subsec-int}

After adding the interaction term $S_{\rm int}$, the structure of the commutators changes.
Since there are no the gauge symmetries \eqref{addsym} anymore, there are no linear combinations of $\psi_X$ which commute with all $\CH_A$.
As a result, all of them give rise to non-trivial stability equations generalizing \eqref{stab-Hpsi}
\be
\left\{\psi_X,\int \de^3 \yc N^A\CH_A\right\}=
M_{XA}N^A=0,
\label{stab-HpsiX}
\ee
where $M_{XA}$ denotes the matrix of commutators.
Its evaluation is discussed in appendix \ref{ap-com} although we refrain from providing explicit expressions
for its entries since they are cumbersome and not illuminating.
The  stability conditions \eqref{stab-HpsiX} represent a system of 6 linear homogeneous equations on 5 Lagrange multipliers $N^A$.
Furthermore, the vanishing solution $N^A=0$ is not physically acceptable.
Therefore, among these 6 equations there should exist at least 2 which are not conditions
on the Lagrange multipliers, but secondary constraints.
In this case the solution necessarily has the form $N^A=\Nt \bfN^A$ where $\bfN^A$ are some functions on the phase space
fixed by the stability conditions.

Unfortunately, we have not been able to find any simple expression for the secondary constraints,
but formally they can be represented as conditions of vanishing determinants. For instance, they can be chosen as
\be
\CS_\pm=\det\{\psi^\pm_{A'}, \CH_B\}=0,
\ee
where $\psi^\pm_{A'}$ denotes the vector of constraints $(\psi_a,\psi_\pm,\psi_0)$.
Then 6 constraints $\psi_X$ form second class pairs with 6 constraints comprising $\CS_+$, $\CS_-$ and 4 constraints out of $(\hCV_a,\CH_+,\CH_-)$,
whereas the total Hamiltonian
\be
\CH=\hbfN^a\hCV_a+\bfN_+\CH_++\bfN_-\CH_-+\cdots
\label{totHam}
\ee
remains first class, as in \eqref{totHampsiS}.

The dimension of the constrained phase space in this case is given by $72-2\times 10-36=16$,
which again corresponds to 8 degrees of freedom having the same interpretation as above.

\section{Foliation-preserving model}
\label{sec-sector}

In this section we present another model which can be viewed as a restriction of the previous one to a particular sector.
The motivation to consider it comes from the observation that under the condition
\be
\chip^i=\chim^i,
\label{eqchi}
\ee
which can also be written in a Lorentz covariant form as
\be
(\eepi{a}\tXm)=0 \ \Leftrightarrow\ (\eemi{a}\tXp)=0,
\label{constr-eqchi}
\ee
the commutators of constraints (see Appendix \ref{ap-com}) significantly simplify.
In fact, these simplifications are not accidental, but reflect a degeneracy of the sector \eqref{eqchi},
manifest already from the fact that the constraints $\psi_a$ defined in \eqref{comp-psia} become identically zero.
Moreover, it is easy to check that the 9 constraints defined by\footnote{There are only 9 independent constraints
among $\psi_a^I$ because they satisfy $X_I\psi_a^I=0$.
Here $X^I$ denotes either $\Xp^I$ or $\Xm^I$ which are equal under the condition \eqref{eqchi}.
Note that this is not true however for $\tXp^I$ or $\tXm^I$.}
\be
\psi_a^I\equiv \psi_a^{IJ}X_J
\label{9psi}
\ee
weakly commute with both $\phi_\pm^{ab}$ and the constraints \eqref{constr-eqchi} which will be denoted by $\Delta_a$.
This implies that in the sector of equal $\chipm^i$, the matrix of commutators of $\psi_a^{IJ}$ and $\phi_\pm^{ab}$
acquires an additional degeneracy and some of the constraints which formed second class pairs
do not do this anymore.

More precisely, one finds that only 9 constraints among $\psi_a^{IJ}$
form such second class pairs with $\phi_\pm^{ab}$ and $\Delta_a$. The other 9 constraints given in \eqref{9psi} remain commuting.
As a consequence, among $\phi_\pm^{ab}$ and $\Delta_a$ there are also 6 constraints commuting with all $\psi_a^{IJ}$.
Explicitly, they can be represented as
\begin{align}\label{defmuij}
& \mu^{ab}=\phi_+^{ab}+\phi_-^{ab}\\\nn
&\quad -\eta_{IK}\eet^{(a}_J \(\qp^{b)c}\eepi{c}^K-\qm^{b)c}\eemi{c}^K\)
\(\sqrt{\frac{\qm}{\qp}}\,\eepi{g}^I\eepi{f}^J\phi_+^{gf}-\sqrt{\frac{\qp}{\qm}}\,\eemi{g}^I\eemi{f}^J\phi_-^{gf}
+{\eps^{IJ}}_{KL}\tXp^K\tXm^L\),
\end{align}
where $\eet^{a}_I$ is the inverse of $\Em\eepi{a}^I+\Ep\eemi{a}^I$, and we notice that the last term is proportional to $\Delta_a$.
Thus, in this sector one gets more constraints which generate non-trivial stability conditions.
There are $9+6+3+2=20$ such constraints: $\psi_a^I$, $\mu^{ab}$, $\hCV_b$ and $\CH_\pm$.

Let us briefly discuss the resulting constraint structure without going into the details of the calculations.
First, the stability of $\psi_a^I$ and $\mu^{ab}$ generates $9+6=15$ equations which are
homogeneous equations on 5 Lagrange multipliers $N^A=(\hN^a,\Ntp,\Ntm)$. Thus, one could think that, as in the previous analysis,
the requirement of non-vanishing of the two lapses guarantees that they fix only 4 Lagrange multipliers and the remaining equations
generate 11 secondary constraints.
However, now there is a new feature: the stability equations involve spatial derivatives of the lapses
which arise from commutators with $\mu^{ab}$ \eqref{defmuij}. In such situation the system of equations is not required
to be fully degenerate to have a non-vanishing solution.
Therefore, the 15 stability conditions fix all 5 Lagrange multipliers up to a constant
and generate only 10 secondary constraints. Altogether 30 constraints form 15 second class pairs
and one remains with $72-2\times(9+15)-2\times 9=6$ dimensional phase space, i.e. 3 degrees of freedom.

The striking difference with the previous cases is that now there are only 9 first class constraints.
This is because one lost time diffeomorphisms since both lapses have been fixed by the stability conditions
turning both Hamiltonian constraints into second class.
This is consistent with the observation that the sector analyzed here can be obtained by adding the constraint $\Delta_a$ directly to the action.
However, the term generating it cannot be written in a spacetime covariant form.
The best one can do is to write
\be
S_{\rm res}[\eep,\eem,\omega]=S[\eep,\eem,\omega]+\int \de^4 x\, \lambda^a_0 \,\eepi{a}^I\eemi{I}^0,
\label{term-constrchi}
\ee
where $\lambda^a_0$ is some Lagrange multiplier field. Such term preserves the Lorentz symmetry and spatial diffeomorphisms,
but breaks time diffeomorphisms. It is this breaking that is responsible for the appearance of the third degree of freedom in this model.

Note however that the additional term in \eqref{term-constrchi} is still invariant under reparametrizations of time which
are independent of spatial coordinates $x^0\to x'{}^0(x^0)$. This means that time diffeomorphisms are not broken completely, but their
global zero mode remains to be a symmetry.
Remarkably, this is consistent with the canonical analysis sketched above which showed that
the lapses are fixed by the stability conditions only up to a constant
which is promoted to a function of time.
These reduced symmetries are precisely consistent with a reduction of full diffeomorphism invariance to foliation-preserving
diffeomorphism invariance.

Thus, the model \eqref{term-constrchi} represents an interesting mild modification of general relativity, in the sense
that there is only one additional degree of freedom, related to the breaking of time diffeomorphisms.

\section{Discussion}
\label{sec-disc}

In this paper we analyzed a bi-gravity model obtained starting from the first order formulation of ghost-free bi-gravity
and imposing that the two tetrads are charged under the same Lorentz connection.
We found that this changes strongly the dynamics and the spectrum of the original theory.
In particular, the linearization around any background with conformally related tetrads reduces to linearized GR and thus propagates
only the 2 degrees of freedom of a massless graviton.
This means that all solutions of linearized Einstein's theory can be included in our model,
as argued at the end of section~\ref{sec-linear}.

At the non-linear level, we found 8 propagating degrees of freedom.
Besides the 2 degrees of freedom of a massless graviton, there are 6 additional modes
which can be traced back to the lost of the symmetricity conditions \Ref{symcon} and thus
to the absence of the equivalence with the metric formalism.
The precise properties of the additional degrees of freedom remain to be investigated.
To that end, one should either compute
higher orders of the perturbative expansion around the doubly-flat background,
or linearize the theory around a background with non-conformally equivalent
tetrads so that these degrees of freedom become visible in the linear approximation.
The problem with the latter approach is that the solution of the connection equation \eqref{domB} becomes much more complicated,
see e.g. \cite{Beke:2011mu},
and one may expect that the geometric interpretation of the corresponding results will be intricate.

While we have no definite results about the nature of the additional degrees of freedom,\footnote{
The number of additional degrees of freedom suggests that they may be carried by the antisymmetric `off-diagonal' variables $b_{[\m\n]}$.}
we were able to identify the mechanism at play in hiding them around conformally related backgrounds: this originates in
the non-linear dependence of their momenta on the off-diagonal tetrad components.
Furthermore, we claim that these additional propagating modes
contain neither the BD ghost, nor a massive graviton.
This may appear at odds with the arguments of  \cite{deRham:2013tfa,deRham:2015cha},
but it is simply a consequence of the fact that the two constraints which remove the BD ghost in ghost-free bi-gravity, are
still present in our model.
At the same time, the absence of a massive graviton is expected because
the additional degrees of freedom originate from 
a different set of canonical fields than the one which gives rise to such graviton in standard bi-gravity.
In our case these are the fields which are typically fixed by the symmetricity constraint \eqref{symcon}.
In our model this constraint is absent and the corresponding fields become dynamical.\footnote{Note that 
degrees of freedom of a similar nature have recently appeared also in a three-dimensional model introduced in \cite{Geiller:2018ain}.}

One may try to compare these results with \cite{deRham:2015rxa}, where an equivalent action was considered,
but taking only one of the tetrads dynamical. In that case the authors argued that the phase space of the theory
should be either $2\times 10$ or $ 2\times 8$-dimensional,
which was interpreted as describing 5 modes of a massive graviton plus 5 or 3 additional modes.
One could then think that the extension to the bi-gravity case just adds the two modes of a massless graviton.
If this was true, our model would have either 10 or 12 degrees of freedom, among which one would find,
in particular, one massless and one massive graviton.
However, such conclusion disagrees with our findings.

It is interesting that our model has 8 degrees of freedom independently of the presence of the interaction term \eqref{Sint}
(which includes the two cosmological constants).
In the case where this term is absent, the model possesses two additional gauge symmetries which however do not change the number of degrees of freedom:
while they do convert two second class constraints into first class, two other second class constraints drop out.

Our model has also an interesting twist presented in section \ref{sec-sector}.
It is obtained by imposing a set of additional constraints which restrict the model into a certain sector degenerate
from the point of view of the canonical structure.
The resulting model turns out to be foliation-preserving and propagates 3 degrees of freedom, a massless graviton and a scalar.
This situation is similar to other foliation-preserving modified theories of gravity such as Einstein-aether theory (with hypersurface-orthogonal vector)
\cite{Jacobson:2008aj}, Horava-Lifshitz gravity \cite{Horava:2009uw},
which attracted much attention, or the recently introduced generalized unimodular gravity \cite{Barvinsky:2019agh}.
These theories have a wide range of applications \cite{Blas:2014aca,Gubitosi:2012hu}, and it
would be interesting to further investigate and compare to the existing literature the model here presented.

There are other possible generalizations of the model. For instance, one can include parity breaking terms
similar to the one which introduces the Immirzi parameter \cite{Holst:1995pc}. With two tetrads, there are three such new terms.
One may also study more general interaction potentials to see whether they lead to the appearance of the BD ghost,
as it happens in the standard bi-gravity. Another possibility is to consider more than two tetrads charged under the same connection.
As argued in the Introduction, we expect that such model should still
contain only one graviton in its spectrum.
And of course, it is crucial to understand what is the correct way to introduce matter couplings.

Coming to applications and further developments, one may hope that this model,
or any of its extensions including the foliation-preserving one, can be useful for cosmology.
The modified dynamics and the additional degrees of freedom
may lead to phenomenologically interesting infrared modifications of
gravity, be it at cosmological or galactical scales.
On the other hand, the fact they are invisible at linearized level may be helpful
in ensuring that the model is not in contradiction with the known experimental results.
Finally, it would also be interesting to understand how our model changes the discussion
of horizon structures and black holes from ordinary bi-gravity \cite{Deffayet:2011rh}.

To conclude, we wish to highlight a possible conceptual lesson of our work --- the crucial role of the connection field in carrying the gravitational
interaction, as opposed to the tetrad (or metric). In fact, one may expect that a theory with two tetrads will
always contain two gravitons, but we have shown otherwise, exposing a theory with two tetrads, a single connection and a single graviton.
This lesson resonates with many approaches to classical and quantum gravity
where one moves the emphasis from the tetrad (or metric) to the connection field (see, for instance,
\cite{Rovelli:2004tv,Freidel:2005ak,Krasnov:2012pd}).

\section*{Acknowledgements}
We would like to thank Lavinia Heisenberg, Federico Piazza and Andrew Tolley for discussions.

\appendix

\section{Details of the linearization}
\label{ap-linear}

In this Appendix we provide additional details and explicit formulas for the linearization considered in section \ref{subsec-conf}.
The equation for the connection perturbations is given by
\be
\label{omEq1}
\eps_{IJKL}\sum_{s=\pm}\( f_{s}^K \w d_{\bar\om} \bar e_{s}^{L} + \bar e_{s}^K \w d_{\bar\om}f_{s}^{L} + \bar e_{s}^K \w w^L{}_M\w \bar e_{s}^{M}\)= 0.
\ee
For generic backgrounds, this presents the same difficulty of the full equation, namely the bivector multiplying the perturbed connection is not simple,
and its inversion is complicated. The situation changes significantly for conformally related backgrounds.
First, the background connection is Levi-Civita with respect to the effective tetrad $\bar\eaux^I=e^\Phi \bar e^I$, as already shown in the main text:
\be\label{lutece}
\bar\om_\mu^{IJ}= \bar\eaux^{\n I}\bar\na_\m \bar\eaux^J_\n.
\ee
Here $\bar\na_\mu$ is the covariant derivative compatible with the effective background metric $\bar\gaux_{\mu\nu}$.
As a consequence, the covariant derivative $D_\m=\p_\m+\om_\m$ is compatible with projecting the internal indices using the effective tetrad, namely
\be
D_\m v^I =\bar\eaux^I_\n\bar\na_\m (\bar\eaux^\n_I v^I).
\ee
In addition, we have the property
\be\label{dT}
d_{\bar\om}\bar e^I_\pm = d(e^{-\Phi}\Om_\pm)\w\bar\eaux^I.
\ee
That is, the on-shell connection is Levi-Civita with respect to the effective tetrad,
but carries torsion with respect to the two fundamental tetrads.
The torsion vanishes only when the backgrounds coincide exactly, like for the doubly-flat background.

Secondly, we can identify the `diagonal' and `off-diagonal' tetrad perturbations
\be
f_{\pm,\m}^I = \f{e^{\Phi}}{2\Om_\pm}(f_\m^I\pm b_\m^I),
\ee
so that they are related to the tensorial perturbations, defined in the same
> way as in \Ref{defcombfluct-flat}, by
\be
f_{\m\n}= f^I_\m \eaux_{I\n} ,
\qquad
b_{\m\n}=b^I_\m \eaux_{I\n}.
\ee
Using \Ref{dT} we can rewrite \eqref{omEq1} in terms of the `diagonal' tetrad perturbations only,
\be\label{omEq2}
\eps_{IJKL}\(\eaux^K\w d_{\bar\om}f^L +\eaux^K\w {w^L}_M\w \eaux^M \)=0.
\ee
This equation can now be recognized as the first order expansion of the standard torsion free condition, and admits the unique Levi-Civita solution
\begin{align}\label{wonshell}
w_\mu^{IJ}= w^{IJ}_\m(\eaux, f^I)
= \bar \eaux^{\n[I}\bar \eaux^{J]\r }\(\bar\na_\r s_{\m\n}+\f12\bar\na_\m a_{\n\r}\).
\end{align}
We remark that even if the tetrad perturbation are arbitrary and not conformally related, the on-shell connection perturbation is again Levi-Civita.
This fact is crucial in recovering the Einstein's equations below.

To see that, we consider now the equations for tetrad perturbations which take the following form
\be\label{EqEin-start}
\hf\,\eps_{IJKL} \teps^{\m\n\r\s} \(\hf\, \fpmi{\nu}^J F^{KL}_{\r\s}(\bar\om) + \bar e_{\pm,\nu}^J D_\r w_\s^{KL} \)
+  \bar\eaux \bar\eaux^{[\mu}_I \bar\gaux^{\rho]\sigma}
\sum_{s=\pm} \frac{c_{\pm s}}{e^{\Phi} \Om_s}\,
(f_{\r\s}+s b_{\r\s})=0,
\ee
where we introduced
\be
\begin{split}
c_{++} = &\, {\beta_2} \Omega_-^2 + 2\beta_3 \Omega_- \Omega_+ + {\beta_4} \Omega_+^2,
\\
c_{+-} =
c_{-+} =&\, {\beta_1} \Omega_-^2 + 2\beta_2 \Omega_- \Omega_+ + {\beta_3} \Omega_+^2,
\\
c_{--} = &\,{\beta_0} \Omega_-^2 + 2\beta_1 \Omega_- \Omega_+ + {\beta_2} \Omega_+^2.
\end{split}
\ee
These coefficients are related to the coefficients $\lambda_\pm$ defined in \eqref{deflampm} by
\be
\begin{split}
\Om_+ c_{++} +\Om_- c_{-+}=&\, \lambda_+,
\\
\Om_+ c_{+-} +\Om_- c_{--}=&\, \lambda_-.
\end{split}
\label{rel-clam}
\ee

Substituting the solution \eqref{omEq1} for $w_\mu^{IJ}$ into \eqref{EqEin-start}
and using the tetrad equations for the background, one finds after some algebra
\be
\begin{split}
& G^{(1)}_{\m\n}[\bar\gaux;s] + \Lambda s_{\m\n}-A_\pm( f_{\n\m}-\gaux_{\n\m}{f_\rho}^\rho)  +Q_\pm (b_{\n\m}-\gaux_{\n\m}{b_\rho}^\rho)
\\
&\qquad +\bar C_{\m\r\n\s}\left[\(1-\f {e^{2\Phi}}{2\Om_\pm^2}\) f^{\r\s}\mp \f {e^{2\Phi}}{2\Om_\pm^2}\,b^{\r\s}\right] = 0,
\end{split}
\label{eqEinst1}
\ee
where $\bar C_{\mu\nu\rho\sigma}$ is the Weyl tensor of the effective metric $\bar\gaux_{\mu\nu}$,
\be
\begin{split}
G^{(1)}_{\m\n}[g;s] =&\,  - \f12\, \na_{\r} \na^\r s_{\m\n} + \na_{(\m} \na_\r {s_{\n)}}^\r
-\f12\, \na_\m\na_\n {s_\rho}^\rho -\f12\, g_{\m\n}( \na_\r\na_\s s^{\r\s} - \na_{\r} \na^\r {s_\s}^\s)
\\ &\,
+ R_{(\m\r}{s_{\n)}}^\r - R_{\m\r\n\s}s^{\r\s} +\f12\, g_{\m\n} R_{\r\s}s^{\r\s} -\f12\, R s_{\m\n}
\end{split}
\label{deflinEinst}
\ee
is the linear term in $s_{\mu\nu}$ of the expansion of the Einstein tensor evaluated on the metric $g_{\mu\nu}+s_{\mu\nu}$,
and we introduced two sets of coefficients
\be
\begin{split}
A_\pm =&\, -\frac{2}{3}\L -\frac{\L}{6}\, \frac{e^{2\Phi}}{\Om_\pm^2} + \frac{1}{2\Om_\pm}\sum_{s=\pm} \f{c_{\pm s}}{\Om_s},
 \\
Q_\pm =&\, \pm\frac{\L}{6}\, \frac{e^{2\Phi}}{\Om_\pm^2} - \frac{1}{2\Om_\pm}\sum_{s=\pm} s\,\frac{c_{\pm s}}{\Om_s}.
\end{split}
\ee
Taking into account the definition of $\Lambda$ \eqref{defLam}, the constraint \eqref{constrOmpm} and the relations \eqref{rel-clam},
it is easy to show that these coefficients satisfy
\beq
\Om_+^2 Q_+ +\Om_-^2 Q_-&=&0,
\label{relQOm}
\\
\Om_+^2 A_+ +\Om_-^2 A_-&=&0.
\label{relAOm}
\eeq
Using these properties, the two equations \eqref{eqEinst1} can be rewritten as
\beq
&& G^{(1)}_{\m\n}[\bar\gaux;s] + \Lambda s_{\m\n}
\label{eqEinst2}\\
&& +Q_\pm \Bigl[(b_{\n\m}-\gaux_{\n\m}{b_\rho}^\rho)-B( f_{\n\m}-\gaux_{\n\m}{f_\rho}^\rho)
-W\bar C_{\m\r\n\s}\((\Om_+^2+\Om_-^2)b^{\r\s}-(\Om_+^2-\Om_-^2) f^{\r\s}\)\Bigr] = 0,
\nn
\eeq
where
\be
B\equiv\frac{A_+}{Q_+}=\frac{A_-}{Q_-},
\qquad
W\equiv \frac{1}{2\Om_+^2Q_+}=-\frac{1}{2\Om_-^2Q_-}.
\label{defW}
\ee
Since $Q_+\ne Q_-$, the equations are equivalent to vanishing of the two lines separately.
The vanishing of the first line coincides with the linearized Einstein equations \eqref{linEinst}, whereas the vanishing of the second
gives the relation \eqref{eq-bf} due to the tracelessness property of the Weyl tensor.
Thus, the `off-diagonal' perturbations $b_{\mu\nu}$ are completely determined by the `diagonal ones',
the symmetric part $s_{\mu\nu}$ of the `diagonal' perturbations satisfies the linearized Einstein equations, whereas
its antisymmetric part $a_{\mu\nu}$ remains unrestricted and hence pure gauge.

\section{The action of the additional gauge symmetries}
\label{ap-sym}

In this appendix we present the action of the gauge symmetries \eqref{addsym}
on the variables appearing in the $3+1$ decomposition \eqref{dec-tetrad}.

For the first symmetry $\CS_1$, one obtains
\be
\begin{split}
\delta \tPpmi{IJ}^a=&\, \pm \frac{\eps}{2}\, \eps_{IJKL}\teps^{abc}\eepi{b}^K\eemi{c}^L,
\\
\delta \tXpm^I=&\, \mp 2\,\eps\, {\tPpm^{aI}}_J \eempi{a}^J,
\\
\delta \Npm^a=&\,  \eps\, \qpm^{ab}\(- (\eepmi{b}\eempi{c}) \hN^c \pm(\eemi{b}\tXp)\Ntp \pm(\eepi{b}\tXm)\Ntm\),
\\
\delta\Ntpm =&\, \eps \(\qpm^{-1}(\eempi{a}\tXpm)\hN^a\mp \qpm^{ab}(\eepi{a}\eemi{b})\Ntpm\mp \qpm^{-1}(\tXp\tXm)\Ntmp \).
\end{split}
\label{transS1N}
\ee

For the second symmetry $\CS_2$, one first finds the transformation of the tetrad
\be
\begin{split}
\delta\eepmi{\mu}^I=\pm \eps\,e_\mp
\[\hf\,\eepmi{\mu}^I \eempi{J}^\nu\eepmi{\nu}^J-\eepmi{\mu}^J \eempi{J}^\nu\eepmi{\nu}^I \] .
\end{split}
\label{transS2}
\ee
Taking into account that in each sector the decomposition \eqref{dec-tetrad} implies the following form of the inverse tetrad
\be
e^0_I=-e^{-1} \tX_I,
\qquad
e^a_I=e^{-1}\(N^a\tX_I-2\Nt\tP^a_{IJ}\tX^J\),
\ee
where the determinant of the tetrad is given by $e=-\Nt \tX^2=\Nt E^2 (1-\chi^2)$, the transformation \eqref{transS2} can be rewritten as
\beq
\delta \eepmi{\mu}^I
&=& \eps\biggl[
-\hf\,\eepmi{\mu}^I\Bigl(\hN^b(\eepmi{b}\tXmp)\pm \Ntpm (\tXp\tXm)\mp \Ntmp \qmp\qmp^{bc}(\eepi{b}\eemi{c})\Bigr)
\biggr.
\nn\\
&& \biggl.
+\eepmi{\mu}^J\Bigl(\hN^b\eepmi{b}^I\tXmpi{J}\pm \Ntpm \tXpm^I\tXmpi{J}\mp \Ntmp \qmp\qmp^{bc}\eepmi{b}^I\eempi{c}^J\Bigr)\biggr].
\eeq
Starting either from this result or directly from \eqref{addgsym2}, one can also derive the following transformations
\be
\begin{split}
\delta \tPpmi{IJ}^a=&\,  \pm\eps\,\eta_{IK}\eta_{JL}
\[\hN^a \tXp^{[K}\tXm^{L]}+\Ntp\qp\qp^{ab}\eepi{b}^{[K}\tXm^{L]}+\Ntm\qm\qm^{ab}\eemi{b}^{[K}\tXp^{L]}\],
\\
\delta \tXpm^I=&\, -\frac{\eps}{2}\, \Bigl[\hN^a(\eepmi{a}\tXmp)\tXpm^I\pm \Ntpm\(\tXmp^I\tXpm^2+\tXpm^I(\tXp\tXm)\)
\Bigr.
\\
&\, \Bigl.
\pm \Ntmp\qmp\qmp^{ab}(\eepi{a}\eemi{b})\tXpm^I\Bigr],
\\
\delta \Npm^a=&\,  \eps\, \Ntpm\[\hN^a(\tXp\tXm)\mp \Ntp\qp\qp^{ab} (\eepi{b}\tXm) \mp \Ntm\qm\qm^{ab} (\eemi{b}\tXp)\],
\\
\delta\Ntpm =&\, \pm 2\,\eps\, \Ntpm^2(\tXp\tXm).
\end{split}
\label{transS2N}
\ee

An important observation is that in both transformations \eqref{transS1N} and \eqref{transS2N} the shifts enter only in the combination $\hN^a$.
This is consistent with the general form of the transformation of the Lagrange multipliers \eqref{transLM} and the form of the constraint algebra.
Another remark is that, in contrast to the first symmetry, the transformations of the canonical variables under the second
are proportional to the Lagrange multipliers, whereas the transformations of the Lagrange multipliers themselves are quadratic in them.
This indicates that the canonical generator of the second symmetry is constructed using these Lagrange multipliers, in agreement with \eqref{genS2}.

\section{Constraints and commutators}
\label{ap-com}

The constraints appearing in the decomposed action \eqref{decompaction} have the following explicit form
\beq
\CG_{IJ}&=&D^{(+)}_a\tPpi{IJ}^a+D^{(-)}_a\tPmi{IJ}^a,
\nn\\
\CV_{\pm,a}&=&-\tPpmi{IJ}^b F_{ab}^{IJ}(\ompm)\pm \mhD_a,
\label{primconstr}
\\
\CH_\pm&=&2\tPpmi{IK}^a{\tPpm^{b,K}}_{\!\! J} F_{ab}^{IJ}(\ompm)+\mHpm,
\nn
\eeq
where $D_a^{(\pm)}$ are the covariant derivatives with respect to $\om_{\pm,a}^{IJ}$, and $\mhD_a$ and $\mHpm$
are contributions of the interaction term \eqref{Sint} given by
\be
\begin{split}
\mhD_a
=&\, \beta_1( \eepi{a}\tXm)-\beta_2 \eps^{IJKL}\epst_{abc}\tPpi{IJ}^b\tPmi{KL}^c-\beta_3(\eemi{a}\tXp),
\\
\mHp
=&\, -\beta_0\qp
+\beta_1(\tXp\tXm)+2\beta_2 \eepi{a}^I\Xp^J\tPmi{IJ}^a -\beta_3 \qp \ddp,
\\
\mHm
=&\,-\beta_1 \qm \ddm+2\beta_2 \eemi{a}^I\Xm^J\tPpi{IJ}^a +\beta_3 (\tXp\tXm)
-\beta_4\qm,
\end{split}
\label{defintconstr}
\ee
where we introduced a convenient notation $\ddpm=\qpm^{ab}(\eepi{a}\eemi{b})$.

Before we proceed, let us note the following useful identities\footnote{We omit the indices $\pm$ distinguishing the two sectors.}
\begin{subequations}\label{identP}
\be
\tP^a_{IJ}=q^{ab}e_b^{[I}\tX^{J]},
\qquad
\tP^a_{IJ} e_a^K= \delta_{[I}^K \tX_{J]},
\qquad
\tP^a_{IJ} e_b^J=-\hf\, \delta^a_b \tX_I,
\qquad
\tP^a_{IJ} \tX^J=-\frac{q}{2}\,q^{ab} e_{b,I},
\label{ident-tPcontr}
\ee
\be
q^{ab} e_a^I e_b^J-q^{-1} \tX^I \tX^J=\eta^{IJ},
\label{ident-compl1}
\ee
\be
\frac{q}{4}\, q^{ac} e_c^I g^{bd} e_d^J-\frac{1}{4}\, q^{ab} \tX^I \tX^J=\tP^a_{IK}{\tP^{b,K}}_J,
\label{ident-compl2}
\ee
\be
\frac14\,\eps_{IJKL}\tX^K e_a^L=\epst_{abc}\tP^b_{IK}{\tP^{c,K}}_J,
\label{ident-PP}
\ee
\be
\frac{1}{2}\, \eps_{IJNL} e_a^L\( \eta^{NK} +q^{-1}\tX^N\tX^K\)
=\epst_{abc}\tP_{IJ}^b q^{cd} e_d^K ,
\label{propr7}
\ee
\end{subequations}
and that the difference of two curvatures vanishes on the surface of the constraints $\psi_a^{IJ}$.
Indeed, one has
\be
F_{ab}^{IJ}(\omp)-F_{ab}^{IJ}(\omm)=\p_a\psi_b^{IJ}-\p_b\psi_a^{IJ}+2\psi_{aK}^{[I}\ompi{b}^{KJ]}+2\ommi{a}^{[IK}{\psi_{bK}}^{J]}.
\ee
This property ensures that in the right hand side of commutators one can always do the replacement
\be
F_{ab}^{IJ}(\ompm)\approx \hf \(F_{ab}^{IJ}(\omp)+F_{ab}^{IJ}(\omm)\)\equiv \CF_{ab}^{IJ}.
\ee

First, we provide commutators between the constraints $\psi_X=(\psi_a, \psi_+,\psi_-,\psi_0)$ \eqref{psiX} and
$\CH_A=(\hCV_a=\hft(\CV_{+,a}-\CV_{-,a}),\CH_+,\CH_-)$ \eqref{primconstr} in the case of vanishing $\beta_k$, $k=0,\dots,4$.
Introducing the relevant part of the Hamiltonian
\be
H=\int \de^3 \xc\(\hN^a \hCV_a+\Ntp\CH_++\Ntm\CH_-\)
\ee
and using the above properties, it is straightforward to obtain
\begin{subequations}\label{comm-psi-constr}
\beq
\{\psi_a,H\}&\approx& -(\tXp\tXm)\(\Ntp\CV_{+,a}+\Ntm\CV_{-,a}\)
\label{psiaH}\\
&&
+\(\hN^b \tXp^I\tXm^J+ \Ntp \qp\qp^{bc}\eepi{c}^I\tXm^J-\Ntm\tXp^I \qm\qm^{bc}\eemi{c}^J\)\eta_{IK}\eta_{JL}\CF_{ab}^{KL},
\nn\\
\{\psi_+,H\}&\approx&\qp\qp^{ab}\eepi{b}^I\tXmi{I}\(\Ntp\CV_{+,a}-\Ntm\CV_{-,a}\)
-2\Ntp (\tXp\tXm)\CH_+
\label{psipH}\\
&&
+\qp\qp^{ac}\eepi{c}^I\(\hN^b\tXm^J-\Ntm \qm\qm^{bd}\eemi{d}^J\)\eta_{IK}\eta_{JL}\CF_{ab}^{KL},
\nn\\
\{\psi_-,H\}&\approx& \qm\qm^{ab}\eemi{b}^I\tXpi{I}\(\Ntp\CV_{+,a}-\Ntm\CV_{-,a}\)
+2\Ntm (\tXp\tXm)\CH_-
\label{psimH}\\
&&
+\qm\qm^{ac}\eemi{c}^I\(\hN^b\tXp^J+\Ntp \qp\qp^{bd}\eepi{d}^J\)\eta_{IK}\eta_{JL}\CF_{ab}^{KL},
\nn\\
\{\psi_0,H\}&\approx& \teps^{abc}\eps_{IJKL}\eepi{b}^K\eemi{c}^L\(\f{\hN^d}2 \CF_{ad}^{IJ}
+2 \(\Ntp\tPpi{JN}^d-\Ntm\tPmi{JN}^d\) \CF_{ad}^{IN}\).
\label{psi0H}
\eeq
\end{subequations}
From this result it follows that the matrix of commutators defined in \eqref{matrixM}
has the structure shown in that equation with the entries given by
\beq
\CA_{ab}&=&\tXpi{I}\tXmi{J}\CF_{ab}^{IJ},
\nn\\
\CB^\pm_a&=&\qpm\qpm^{bc}\eepmi{c}^I\tXmpi{J}\eta_{IK}\CF_{ab}^{KJ},
\label{defABC}\\
\CC &=&\qp\qp^{ac}\eepi{c}^I \,\qm\qm^{bd}\eemi{d}^J\,\eta_{IK}\eta_{JL}\CF_{ab}^{KL}.
\nn
\eeq

At the same time, one can prove that the commutator \eqref{psi0H} is actually weakly vanishing.
Indeed, the first term can be rewritten as
\beq
\left\{\psi_0,\int \de^3\xc \hN^a \hCV_a\right\}
&\approx & \frac14\, \hN^d\teps^{abc}\eps_{IJKL}\(\eemi{d}^K\eepi{c}^L+\eepi{d}^K\eemi{c}^L\)\CF_{ab}^{IJ}
\nn\\
&=& \frac14\, \hN^d\teps^{abc}\eps_{IJKL}\Bigl(\qp^{gf}(\eepi{g}\eemi{d})\eepi{f}^K\eepi{c}^L+
\qm^{gf}(\eemi{g}\eepi{d})\eemi{f}^K\eemi{c}^L
\Bigr.
\nn\\
&&\Bigl.
-\qp^{-1}(\eemi{d}\tXp)\tXp^K\eepi{c}^L-\qm^{-1}(\eepi{d}\tXm)\tXm^K\eemi{c}^L\Bigr)\CF_{ab}^{IJ}
\nn\\
&=& -\frac14\, \hN^d\eps_{IJKL}\Bigl(\teps^{bcf}\(\qp^{ag}(\eepi{g}\eemi{d})\eepi{c}^K\eepi{f}^L+
\qm^{ag}(\eemi{g}\eepi{d})\eemi{c}^K\eemi{f}^L\)
\Bigr.
\nn\\
&&\Bigl.
+\teps^{abc}\(\qp^{-1}(\eemi{d}\tXp)\tXp^K\eepi{c}^L+\qm^{-1}(\eepi{d},\tXm)\tXm^K\eemi{c}^L\)\Bigr)\CF_{ab}^{IJ}
\nn\\
&=&
- \hN^d\Bigl(\qp^{ag}(\eepi{g}\eemi{d})\tPpi{IJ}^b+\qm^{ag}(\eemi{g}\eepi{d})\tPmi{IJ}^b
\Bigr.
\nn\\
&&\Bigl.
+2\qp^{-1}(\eemi{d}\tXp)\tPpi{IK}^a{\tPp^{bK}}_J+2\qm^{-1}(\eepi{d}\tXm)\tPmi{IK}^a{\tPm^{bK}}_J
\Bigr)\CF_{ab}^{IJ}
\nn\\
&\approx &
\qp^{ab}(\eepi{b}\eemi{c})\hN^c\hCV_{+,a}+\qm^{ab}(\eemi{b}\eepi{c})\hN^c\hCV_{-,a}
\nn\\
&&
-\qp^{-1}\hN^a(\eemi{a}\tXp)\CH_+-\qm^{-1}\hN^a(\eepi{a}\tXm)\CH_-,
\eeq
where at the second step we used the identity \eqref{ident-compl1} and at the fourth step we used
the definition of $\tPpmi{IJ}^a$ \eqref{deftP} and \eqref{ident-PP}.
The second term can be manipulated as follows
\beq
\left\{\psi_0,\int \de^3\xc \Nt_+ \CH_+\right\}
&\approx&\hf\, \Ntp\teps^{abc}\teps^{dgf}\eepi{g}^I\(\qpi{bf}\eemi{c}^J-\eepi{b}^J(\eepi{f}\eemi{c})\)\CF_{ad}^{IJ}
\nn\\
&=&\Ntp\qp\qp^{ac}\eepi{c}^I\qp^{bd}\(\eemi{d}^J -\eepi{g}^J\qp^{gf}(\eepi{f}\eemi{d})+\eepi{d}^J\qp^{gf}(\eepi{g}\eemi{f})
\right.
\nn\\
&&\left.
-(\eepi{d}\eemi{g})\qp^{gf}\eepi{f}^J
\)\eta_{IK}\eta_{JL}\CF_{ab}^{KL}
\nn\\
&=& \Ntp\Bigl(-\qp^{ac}\eepi{c}^I\qp^{bd}\tXp^J(\tXp\eemi{d})+4\ddp\tPp^{aIK}\tPpi{KJ}^b
\Bigr.
\nn\\
&&\Bigl.
-4\tPp^{aIK}\tPpi{KJ}^g\qp^{bd}(\eepi{d}\eemi{g})\Bigr)\eta_{IN}\CF_{ab}^{NJ}
\nn\\
&\approx & -\Ntp(\tXp\eemi{a})\qp^{ab}\CV_{+,b}+2\ddp\Ntp\CH_+
\nn\\
&&
-\hf\,\Ntp\teps^{agf}\eps_{IJKL}\qp^{bc}(\eepi{c}\eemi{g})\tXp^K\eepi{f}^L\CF_{ab}^{IJ},
\eeq
where at the first step we used the definition of $\tPpi{IJ}^a$ and contracted the two $\eps$-factors,
at the third step we used \eqref{ident-compl1} and \eqref{ident-compl2},
and at the last step we applied \eqref{ident-PP} in the last term.
This last term can then be rewritten as
\beq
&&
-\hf\,\teps^{agf}\eps_{IJKL}\qp^{bc}(\eepi{c}\eemi{g})\tXp^K\eepi{f}^L\CF_{ab}^{IJ} =
-\frac14\, \teps^{abc}\eps_{IJKL}\qp^{gf}\((\eepi{g}\eemi{f})\eepi{c}^L-(\eepi{g}\eemi{c})\eepi{f}^L\)\tXp^K\CF_{ab}^{IJ}
\nn\\
&=& -2\ddp\tPpi{IK}^a{\tPp^{bK}}_J\CF_{ab}^{IJ}+\frac14\, \teps^{abc}\eps_{IJKL}\tXp^K\eemi{c}^L \CF_{ab}^{IJ}
\nn\\
&\approx& -\ddp\CH_+
-\frac14\,\teps^{abc}\(2\epst_{cgf}\qm^{gd}(\eemi{d}\tXp)\tPmi{IJ}^f+\qm^{-1}(\tXp\tXm)\eps_{IJKL}\tXm^K\eemi{c}^L\) \CF_{ab}^{IJ}
\nn\\
&\approx& -\ddp\CH_+ +(\tXp\eemi{a})\qm^{ab}\CV_{-,b}
-\qm^{-1}(\tXp\tXm)\CH_-,
\eeq
where at the second step we again used \eqref{ident-PP} and \eqref{ident-compl1}, and at the third step \eqref{propr7}.
Thus, the full commutator reads
\be
\begin{split}
\left\{\psi_0,\int \de^3\xc \Nt_+ \CH_+\right\}\approx & \, -\Ntp(\tXp\eemi{a})\qp^{ab}\CV_{+,b}+\Ntp(\tXp\eemi{a})\qm^{ab}\CV_{-,b}
\\
&\, +\Ntp\ddp\CH_+ -\Ntp\qm^{-1}(\tXp\tXm)\CH_-.
\end{split}
\ee
A similar formula holds for the commutator with $\hCH_-$. It can be obtained by exchanging $(+ \leftrightarrow -)$ and flipping the overall sign.
As a result, all commutators with $\psi_0$ weakly vanish confirming that this is a first class constraint.
Furthermore, the obtained results are perfectly consistent with the transformations of the Lagrange multipliers \eqref{transS1N}
and the general formula \eqref{transLM}.

To include contributions from the interaction terms, one has to first extend the symplectic structure
to the variables $\tXpm^I$ and $\eepmi{a}^I$ which enter the contributions \eqref{defintconstr}, but cannot be expressed
through the canonical variables $\tPpmi{IJ}^a$ in a covariant way.
One way to proceed is to drop the covariance and decompose the connections as  \cite{Alexandrov:1998cu}
\be
\begin{split}
\ompmi{a}^{ij}=&\,\eps^{ijk}\rpmi{kl}\Etpmi{a}^l +\Etpmi{a}^{[i}\ompm^{j]},
\\
\ompmi{a}^{0i}= &\,\etapmi{a}^i-\ompmi{a}^{ij}\chipmi{j},
\end{split}
\label{decomp-ompm}
\ee
so that $\etapmi{a}^i$ and $\ompm^{i}$ become canonically conjugate to $\tEpmi{i}^a$ and $\chipm^i$, respectively,
whereas $\rpm^{ij}$ have vanishing conjugate momenta (cf. \eqref{Skin4}).
Such formalism is equivalent to solving explicitly the simplicity constraints \eqref{conphi} and changing the symplectic structure
to the one given by the corresponding Dirac bracket.\footnote{In this formalism
the distinguishing feature of the constraints \eqref{psiX} is that they are such combinations of $\psi_a^{IJ}$ that are independent of $\rpm^{ij}$.}
However, the lost of covariance makes all computations extremely cumbersome.
Fortunately, it is possible to represent the resulting Dirac brackets in a covariant form \cite{Alexandrov:2013rxa}
\be
\begin{split}
\{\ompmi{a}^{IJ}, \tXpm^K\}_D =&\, -\eta^{K[I} \eepmi{a}^{J]},
\\
\{\ompmi{a}^{IJ}, \eepmi{b}^K\}_D =&\, -2\(\Ptpmi{b}^{IJ}\eepmi{a}^K+\Ptpmi{b}^{K[I} \eepmi{a}^{J]}\).
\end{split}
\label{comm-om-eX}
\ee
But then the simple canonical commutation relations \eqref{canPB} are also replaced by the Dirac bracket
and become more complicated \cite{Alexandrov:2000jw}, which again makes computations quite involved.

Remarkably, in our case it is possible to apply \eqref{comm-om-eX} and still use the canonical form \eqref{canPB}
for the commutator of the connections with $\tPpmi{IJ}^a$, which in particular implies that the results \eqref{comm-psi-constr}
for the part of the commutators without the interaction terms are still valid.
The reason for this is that the relevant stability conditions only require to compute $\{\psi_X,\CH_A\}_D$.
But $\psi_X$ by definition weakly commute with the simplicity constraints $\phi_\pm^{ab}$ which reduces the Dirac bracket to the usual Poisson bracket.

In this way one can compute all relevant commutation relations, $\{\psi_X,\mhD_a\}$ and $\{\psi_X,\mHpm\}$.
However, the resulting expressions are somewhat messy and not illuminating.
In particular, we have not been able to identify any simple form of the secondary constraints which they are supposed to generate.
Due to this reason, we do not provide them in this paper.


\end{document}